\documentclass[twocolumn]{aastex63}

\usepackage{amsmath,amstext}
\usepackage[T1]{fontenc}
\usepackage[figure,figure*]{hypcap}
\usepackage{multirow}
\usepackage[T1]{fontenc} 
\usepackage[utf8]{inputenc} 
\usepackage{float}
\newcommand{\ba}{\begin{eqnarray}}
\newcommand{\ea}{\end{eqnarray}}

\newcommand{\unit}[1]{\ensuremath{\, \mathrm{#1}}}
\newcommand{\respsi}{$\psi=95_{-8}^{+9\:\circ}$} 
\newcommand{\respsival}{$95_{-8}^{+9}$} 
\newcommand{\reslambda}{$\lambda=98_{-12}^{+15\:\circ}$} 
\newcommand{\reslambdaval}{$98_{-12}^{+15}$}  
\newcommand{\resvsini}{$v \sin i = 0.85_{-0.33}^{+0.27} \unit{km/s}$}  
\newcommand{\resvsinival}{$0.85_{-0.33}^{+0.27}$} 
\newcommand{\resistarval}{$46_{-20}^{+27}$} 
\newcommand{\resistar}{$i_\star = 46_{-20}^{+27\:\circ}$}  
\newcommand{\resK}{$8.03_{-0.37}^{+0.38} \unit{m/s}$} 

\newcommand{\PSUAA}{Department of Astronomy \& Astrophysics, 525 Davey Laboratory, The Pennsylvania State University, University Park, PA, 16802, USA}
\newcommand{\PSUCEHW}{Center for Exoplanets and Habitable Worlds, 525 Davey Laboratory, The Pennsylvania State University, University Park, PA, 16802, USA}
\newcommand{\PSETI}{Penn State Extraterrestrial Intelligence Center, 525 Davey Laboratory, The Pennsylvania State University, University Park, PA, 16802, USA}
\newcommand{\UA}{Steward Observatory, The University of Arizona, 933 N.\ Cherry Ave, Tucson, AZ 85721, USA}

\newcommand{\Penn}{Department of Physics and Astronomy, University of Pennsylvania, 209 S 33rd St, Philadelphia, PA 19104, USA}

\newcommand{\STScI}{Space Telescope Science Institute, 3700 San Martin Dr, Baltimore, MD 21218, USA}
\newcommand{\JHU}{Department of Physics and Astronomy, Johns Hopkins University, 3400 N Charles St, Baltimore, MD 21218, USA}
\newcommand{\GoddardESAL}{Exoplanets and Stellar Astrophysics Laboratory, NASA Goddard Space Flight Center, Greenbelt, MD 20771, USA}

\newcommand{\NOAO}{NSF's National Optical-Infrared Astronomy Research Laboratory, 950 N.\ Cherry Ave., Tucson, AZ 85719, USA}

\newcommand{\Macquarie}{Department of Physics and Astronomy, Macquarie University, Balaclava Road, North Ryde, NSW 2109, Australia }
\newcommand{\NIST}{National Institute of Standards \& Technology, 325 Broadway, Boulder, CO 80305, USA}
\newcommand{\CUBoulder}{Department of Physics, 390 UCB, University of Colorado, Boulder, CO 80309, USA}
\newcommand{\JPL}{Jet Propulsion Laboratory, California Institute of Technology, 4800 Oak Grove Drive, Pasadena, California 91109}

\newcommand{\UCI}{Department of Physics \& Astronomy, The University of California, Irvine, Irvine, CA 92697, USA}
\newcommand{\Carleton}{Carleton College, One North College St., Northfield, MN 55057, USA}
\newcommand{\PSUICS}{Institute for Computational and Data Sciences, The Pennsylvania State University, University Park, PA, 16802, USA}
\newcommand{\PSUCASt}{Center for Astrostatistics, 525 Davey Laboratory, The Pennsylvania State University, University Park, PA, 16802, USA}

\newcommand{\Princeton}{Department of Astrophysical Sciences, Princeton University, 4 Ivy Lane, Princeton, NJ 08540, USA}
\newcommand{\RUSSELL}{Henry Norris Russell Fellow}

\newcommand{\Oklahoma}{Homer L. Dodge Department of Physics and Astronomy, University of Oklahoma, 440 W. Brooks Street, Norman, OK 73019, USA}
\newcommand{\CFA}{Harvard-Smithsonian Center for Astrophysics, 60 Garden Street, Cambridge, MA 02138, USA}
\newcommand{\Warwick}{Department of Physics, University of Warwick, Gibbet Hill Road, Coventry CV4 7AL, United Kingdom}
\newcommand{\HWS}{Department of Physics, Hobart and William Smith Colleges, 300 Pulteney Street, Geneva, NY 14456, USA}
\newcommand{\ABIOLCENTER}{Astrobiology Center, 2-21-1 Osawa, Mitaka, Tokyo 181-8588, Japan}
\newcommand{\NAOJ}{National Astronomical Observatory of Japan, NINS, 2-21-1 Osawa, Mitaka, Tokyo 181-8588, Japan}
\newcommand{\UTAustin}{Department of Astronomy, The University of Texas at Austin, 2515 Speedway, Austin, TX 78712, USA}
\newcommand{\MCDonald}{McDonald Observatory and Department of Astronomy, The University of Texas at Austin, 2515 Speedway, Austin, TX 78712, USA}
\newcommand{\UTSpace}{Center for Planetary Systems Habitability, The University of Texas at Austin, 2515 Speedway, Austin, TX 78712, USA}

\shorttitle{GJ\,3470b has a Polar Orbit}
\shortauthors{Stefánsson et al.}

\begin{document}

\title{The Warm Neptune GJ\,3470b has a Polar Orbit}

\correspondingauthor{Guðmundur Stefánsson}
\email{gstefansson@astro.princeton.edu}

\author[0000-0001-7409-5688]{Guðmundur Stefánsson} 
\affil{\Princeton}
\affil{\RUSSELL}

\author[0000-0001-9596-7983]{Suvrath Mahadevan}
\affil{\PSUAA}
\affil{\PSUCEHW}

\author[0000-0003-0412-9314]{Cristobal Petrovich}
\affiliation{Instituto de Astrofísica, Pontificia Universidad Católica de Chile, Av. Vicuña Mackenna 4860, 782-0436 Macul, Santiago, Chile}
\affiliation{Millennium Institute for Astrophysics, Chile}

\author[0000-0002-4265-047X]{Joshua N.\ Winn}
\affil{\Princeton}

\author[0000-0001-8401-4300]{Shubham Kanodia}
\affil{\PSUAA}
\affil{\PSUCEHW}

\author[0000-0003-3130-2282]{Sarah C. Millholland}
\affil{NASA Sagan Fellow}
\affil{\Princeton}

\author[0000-0001-8222-9586]{Marissa Maney}
\affil{\CFA}

\author[0000-0003-4835-0619]{Caleb I. Ca\~nas}
\affil{NASA Earth and Space Science Fellow}
\affil{\PSUAA}
\affil{\PSUCEHW}

\author[0000-0001-9209-1808]{John Wisniewski}
\affil{\Oklahoma}

\author[0000-0003-0149-9678]{Paul Robertson}
\affil{\UCI}

\author[0000-0001-8720-5612]{Joe P.\ Ninan}
\affil{\PSUAA}
\affil{\PSUCEHW}

\author[0000-0001-6545-639X]{Eric B.\ Ford}
\affil{\PSUAA}
\affil{\PSUICS}
\affil{\PSUCEHW}
\affil{\PSUCASt}

\author[0000-0003-4384-7220]{Chad F.\ Bender}
\affil{\UA}

\author[0000-0002-6096-1749]{Cullen H.\ Blake}
\affil{\Penn}

\author[0000-0001-8934-7315]{Heather Cegla}
\affil{\Warwick}

\author[0000-0001-9662-3496]{William D. Cochran}
\affil{\MCDonald}
\affil{\UTSpace}

\author[0000-0002-2144-0764]{Scott A.\ Diddams}
\affil{\NIST}
\affil{\CUBoulder}

\author[0000-0002-3610-6953]{Jiayin Dong}
\affil{\PSUAA}
\affil{\PSUCEHW}

\author[0000-0002-7714-6310]{Michael Endl}
\affil{\MCDonald}
\affil{\UTSpace}

\author[0000-0002-0560-1433]{Connor Fredrick}
\affil{Time and Frequency Division, National Institute of Standards and Technology, 325 Broadway, Boulder, CO 80305, USA}
\affil{Department of Physics, University of Colorado, 2000 Colorado Avenue, Boulder, CO 80309, USA}

\author[0000-0003-1312-9391]{Samuel Halverson}
\affil{\JPL}

\author[0000-0002-1664-3102]{Fred Hearty}
\affil{\PSUAA}
\affil{\PSUCEHW}

\author[0000-0003-1263-8637]{Leslie Hebb}
\affil{\HWS}

\author[0000-0003-3618-7535]{Teruyuki Hirano}
\affil{\ABIOLCENTER}
\affil{\NAOJ}

\author[0000-0002-9082-6337]{Andrea S.J.\ Lin}
\affil{\PSUAA}
\affil{\PSUCEHW}

\author[0000-0002-9632-9382]{Sarah E.\ Logsdon}
\affil{\NOAO}

\author[0000-0003-0790-7492]{Emily Lubar}
\affil{\UTAustin}

\author[0000-0003-0241-8956]{Michael W.\ McElwain}
\affil{\GoddardESAL} 

\author[0000-0001-5000-1018]{Andrew J. Metcalf}
\affiliation{Space Vehicles Directorate, Air Force Research Laboratory, 3550 Aberdeen Ave. SE, Kirtland AFB, NM 87117, USA}
\affiliation{Time and Frequency Division, National Institute of Standards and Technology, 325 Broadway, Boulder, CO 80305, USA} 
\affiliation{Department of Physics, University of Colorado, 2000 Colorado Avenue, Boulder, CO 80309, USA}

\author[0000-0002-0048-2586]{Andrew Monson}
\affil{\PSUAA}
\affil{\PSUCEHW}

\author[0000-0002-2488-7123]{Jayadev Rajagopal}
\affil{\NOAO}

\author[0000-0002-4289-7958]{Lawrence W. Ramsey}
\affil{\PSUAA}
\affil{\PSUCEHW}

\author[0000-0001-8127-5775]{Arpita Roy}
\affil{\STScI}
\affil{\JHU}

\author[0000-0002-4046-987X]{Christian Schwab}
\affil{\Macquarie}

\author[0000-0001-9580-4869]{Heidi Schweiker}
\affil{\NOAO}

\author[0000-0002-4788-8858]{Ryan C. Terrien}
\affil{\Carleton}

\author[0000-0001-6160-5888]{Jason T.\ Wright}
\affil{\PSUAA}
\affil{\PSUCEHW}
\affil{\PSETI}

\begin{abstract}
The warm Neptune GJ\,3470b transits a nearby ($d=29 \unit{pc}$) bright slowly rotating M1.5-dwarf star. Using spectroscopic observations during two transits with the newly commissioned NEID spectrometer on the WIYN 3.5m Telescope at Kitt Peak Observatory, we model the classical Rossiter-Mclaughlin effect yielding a sky-projected obliquity of $\lambda=98_{-12}^{+15\:\circ}$ and a $v \sin i = 0.85_{-0.33}^{+0.27} \unit{km/s}$. Leveraging information about the rotation period and size of the host star, our analysis yields a true obliquity of $\psi=95_{-8}^{+9\:\circ}$, revealing that GJ\,3470b is on a polar orbit. Using radial velocities from HIRES, HARPS and the Habitable-zone Planet Finder, we show that the data are compatible with a long-term RV slope of $\dot{\gamma} = -0.0022 \pm 0.0011 \unit{m/s/day}$ over a baseline of 12.9 years. If the RV slope is due to acceleration from another companion in the system, we show that such a companion is capable of explaining the polar and mildly eccentric orbit of GJ 3470b using two different secular excitation models. The existence of an outer companion can be further constrained with additional RV observations, Gaia astrometry, and future high-contrast imaging observations. Lastly, we show that tidal heating from GJ 3470b's mild eccentricity has most likely inflated the radius of GJ 3470b by a factor of $\sim$1.5-1.7, which could help account for its evaporating atmosphere. 
\end{abstract}

\keywords{Exoplanets -- Transits -- M-dwarf -- Radial Velocity}

\section{Introduction}
\label{sec:intro}
The stellar obliquity ($\psi$), the angle between the stellar spin axis and a planet's orbital axis, is an important parameter of an exoplanet system. Although the planets in the solar system are observed to be well-aligned to the spin axis of the Sun (within $7^\circ$), exoplanetary systems show a broad range of obliquities, ranging from well-aligned to severely misaligned. These results have been interpreted as clues to their formation \citep{albrecht2012}. Different mechanisms have been proposed to explain the tilting of planetary orbits, including primordial misalignment between the star and the protoplanetary disk \citep{lai2011,batygin2012}, nodal precession induced by an inclined companion \citep{yee2018}, the Von Zeipel-Lidov-Kozai meachanism \citep{fabrycky2007,naoz2016,ito2019}, planet-planet scattering \citep{rasio1996,chatterjee2008}, and secular resonance crossings due to a disappearing disk and a massive outer planetary companion \citep{petrovich2020}.

Stellar obliquities can be constrained by exploiting the Rossiter-McLaughlin (RM) effect, the alteration of the rotational broadening kernel of the star's absorption line profiles that occurs during a planetary transit. The RM effect is often observed as a radial velocity anomaly \citep{triaud2018}, and has been observed for hundreds of planetary systems. However, the RM effect is primarily sensitive to the \textit{sky-projected} obliquity ($\lambda$), the angle between the sky projections of the stellar rotation axis and the planet orbital axis\footnote{In the special case where the differential rotation is known or can be measured, the RM and the Reloaded RM techniques can place a constraint on the three-dimensional obliquity \citep[see e.g.,][]{gaudi2007,cegla2016,sasaki2021}.}. To obtain the three-dimensional obliquity $\psi$, observations of the RM effect generally need to be supplemented with a constraint on the inclination $i_\star$ of the stellar rotation axis with respect to the line of sight.

Using a sample of true obliquities $\psi$, \cite{albrecht2021} found evidence that misaligned systems show a preference for nearly polar orbits ($\psi = 80-125^\circ$) rather than spanning the full range of possible obliquities. Most of the available sample consists of hot Jupiters because they allow for the most straightforward measurements. However, hot Jupiters are intrinsically rare \citep{dawson2018}, and it is unclear if planetary systems hosting smaller planets show the same orbital architectures as hot Jupiters. Among the systems studied by \cite{albrecht2021}, a few warm Neptunes ($a/R_\star \gtrsim 8$) orbiting cool stars ($T_{\mathrm{eff}} < 6100 \unit{K}$) have been observed to have polar orbits, including HAT-P-11b \citep{sanchisojeda2011}, GJ\,436b \citep{bourrier2018,bourrier2022}, HD 3167c \citep{dalal2019,bourrier2021}, and WASP-107b \citep{DaiWinn2017,rubenzahl2021}. For three of these planets---HAT-P-11b \citep{allart2018}, GJ 436b \citep{kulow2014,ehrenreich2015}, and WASP-107b \citep{allart2019}---there is evidence for ongoing atmospheric mass loss. Furthermore, two of the planets are known to have outer planetary companions (HAT-P-11b, \citealt{yee2018}; and WASP-107b, \citealt{piaulet2021}) suggesting that they arrived at their current polar orbits through dynamical interactions.

Here we present observations of the RM effect of the low-density warm Neptune GJ\,3470b, which orbits a bright ($V=12.3$, $J=8.8$) M1.5 dwarf star located 29 parsecs away \citep{bonfils2012}. GJ 3470b is known to be undergoing substantial mass loss \citep[e.g.,][]{bourrier2018b,ninan2019gj3470}. We performed the precise radial velocity (RV) observations with the recently commissioned NEID spectrograph \citep{schwab2016} on the WIYN 3.5m Telescope at Kitt Peak Observatory\footnote{The WIYN Observatory is a joint facility of the NSF's National Optical-Infrared Astronomy Research Laboratory, Indiana University, the University of Wisconsin-Madison, Pennsylvania State University, the University of Missouri, the University of California-Irvine, and Purdue University.}. Two transits with NEID reveal that GJ\,3470b has an RM signal consistent with a polar orbit. Additionally, we detect evidence for a long-term acceleration based on RVs reported in the literature and newly obtained with the Habitable-zone Planet Finder, suggesting the existence of an outer companion in the system. With these measurements, GJ\,3470b joins a growing sample of warm Neptunes on polar orbits that are observed to have evaporating atmospheres, suggesting that such systems might share a common formation history involving dynamical interactions with an outer companions in the system \citep[e.g.,][]{bourrier2018,owen2018,correia2020,attia2021}.

\section{Stellar Parameters}
\label{sec:stellar}
Table \ref{tab:stellarparam} lists the stellar parameters used in this work. To obtain precise estimates of the stellar mass and radius, we performed an SED fit of available literature magnitudes of the star using \texttt{EXOFASTv2} \citep{eastman2019}, along with a precise parallax estimate from \texttt{Gaia}. The resulting values agree with the values of the stellar radius ($R=0.48\pm0.04\,R_\odot$) and mass ($M=0.51\pm0.06\,R_\odot$) reported by \cite{biddle2014}. 

The stellar rotation period is particularly important for this RM analysis given the low $v \sin i$ we measure of \resvsini\ to calculate the expected equatorial velocity. The rotation period of the star was measured using several methods. \cite{biddle2014} used photometric observations with the 0.36m Automated Imaging Telescope (AIT) at Fairborn Observatory in Arizona between December 2012 and May 2013, which showed photometric modulations with a period of $20.7 \pm 0.15$ days and an amplitude of 0.01~mag. \cite{kosiarek2019} analyzed additional AIT observations extending to May 2017, confirming the previous measurement and deriving a period of $21.54 \pm 0.49 \unit{days}$ from the entire dataset. Further, \cite{kosiarek2019} saw a corresponding peak in the periodograms of precise radial velocity observations of GJ\,3470. We confirm this signal in the RV residuals in Section \ref{sec:rv}. We adopt a stellar rotation period value of $21.54 \pm 0.49 \unit{days}$ for our analysis, as this value is both seen in the RV residuals discussed in Section \ref{sec:rv} and in the long-baseline photometry in \cite{kosiarek2019}.

\begin{deluxetable}{llcc}
\tablecaption{Summary of stellar parameters used in this work. \label{tab:stellarparam}}
\tabletypesize{\scriptsize}
\tablehead{\colhead{Parameter}       &  \colhead{Description}                                  & \colhead{Value}                                     & \colhead{Reference}}
\startdata
$M_*$                                &  Mass                                                   & $0.527_{-0.026}^{+0.024} \unit{M_{\odot}}$          & (1)        \\
$R_*$                                &  Radius                                                 & $0.500_{-0.016}^{+0.017} \unit{R_{\odot}}$          & (1)        \\
$T_{\mathrm{eff}}$                   &  Effective Temperature                                  & $3622_{-55}^{+58} \unit{K}$                         & (1)        \\
$d$                                  &  Distance                                               & $29.326\pm0.022 \unit{pc}$                          & (2)        \\
Age                                  &  Age                                                    & $0.3-3 \unit{Gyr}$                                  & (3)        \\
$\mathrm{[Fe/H]}$                    &  Metallicity                                            & $0.18 \pm 0.08$                                     & (4)        \\
$P_{\mathrm{rot}}$                   &  Rotation Period                                        & $21.54\pm0.49 \unit{days}$                          & (5)        \\
\enddata
\tablenotetext{}{References are: (1) This work, (2) \cite{bailer-jones2018}, (3) \cite{bonfils2012}, (4) \cite{biddle2014}, (5) \cite{kosiarek2019}.}
\end{deluxetable}

\section{Observations}
\label{sec:obs}

\subsection{Transit Spectroscopy with NEID}
We observed two transits with the NEID spectrograph \citep{schwab2016} on the WIYN 3.5m Telescope at Kitt Peak Observatory, on the nights of 2021 January 1 (2 January UT), and 2021 January 11 (12 January UT). NEID is an actively environmentally stabilized \citep{stefansson2016,robertson2019} fiber-fed \citep{kanodia2018fiber} spectrograph covering the wavelength range from 380 to 930\,nm at a resolving power of $R\approx 110{,}000$ \citep{halverson2016}. We obtained 25 and 24 spectra for the two transits respectively, using an exposure time of 600\,sec. The first night had clear sky conditions and light winds with a median seeing of $0.7\arcsec$. The second night had poorer conditions with high winds of $\sim$20-25mph with poor seeing ranging from 1.5 to 2.5$\arcsec$. This resulted in median signal-to-noise ratio (SNR) on the two nights of 15.2 and 5.4, respectively, evaluated per 1D extracted pixel at a wavelength of $550 \unit{nm}$, and median RV uncertainties of $1.7 \unit{m/s}$ and $4.0 \unit{m/s}$, respectively.

\begin{figure*}[t!]
\begin{center}
\includegraphics[width=\textwidth]{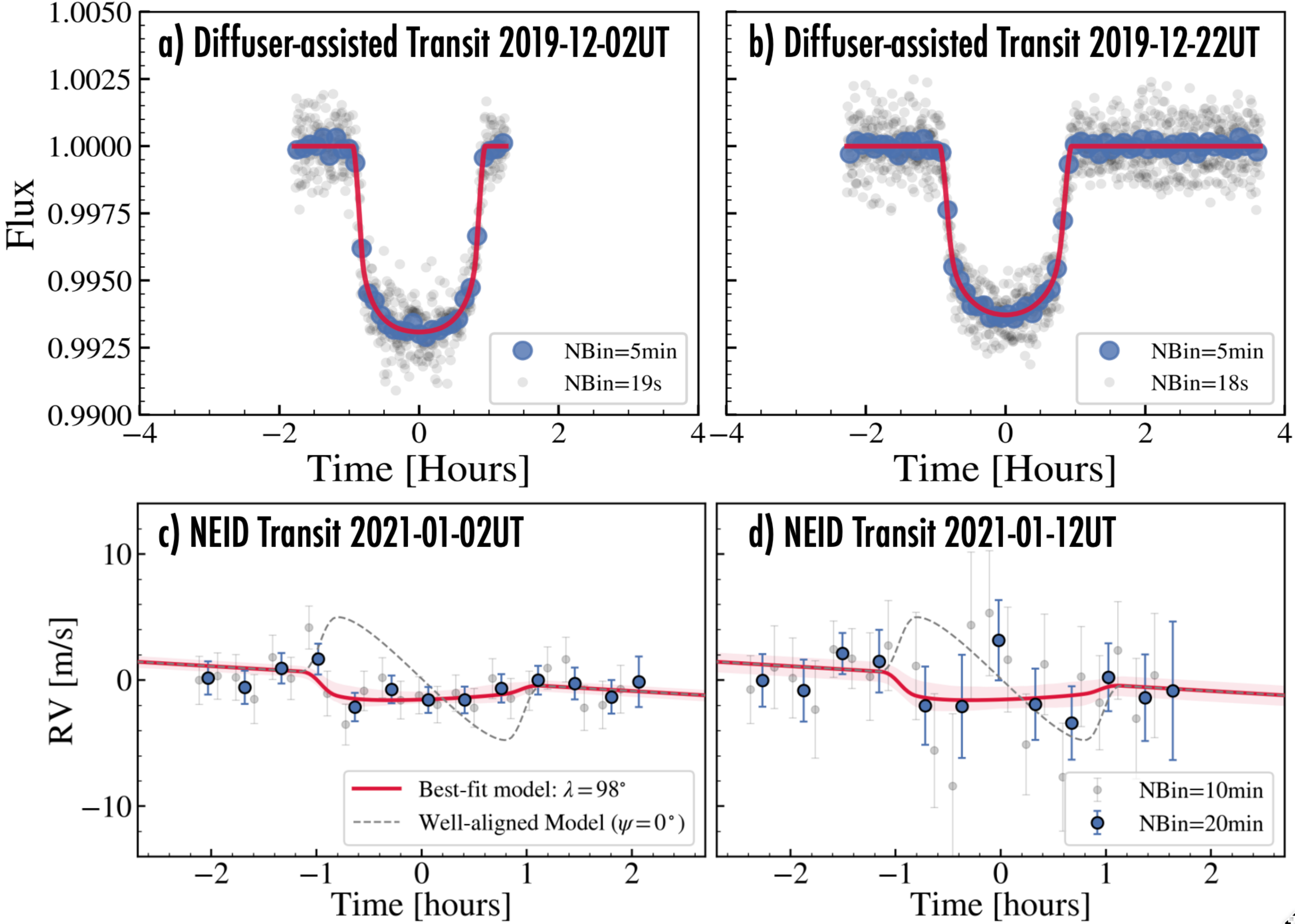}
\vspace{-0.8cm}
\end{center}
\caption{\textbf{Top panels:} Diffuser-assisted transit observations with the ARCTIC imager on the ARC 3.5m Telescope at Apache Point Observatory on (a) 2 December 2019, and (b) 22 December 2019. The best fit model is shown in red, unbinned data are shown in black, and 5-min binned data are shown with the blue points. \textbf{Lower panels:} RM effect of GJ\,3470b as observed by NEID on c) 2 January 2021 UT, and d) 12 January 2021 UT. The best-fit joint model of both RM observations (red) results in a sky-projected obliquity of \reslambda. The red regions show the $1\sigma$ credible intervals. The expected well-aligned model is shown with the grey-dashed lines, which assumes $\psi = 0^\circ$, i.e., that $v \sin i = v_{\mathrm{eq}}$, and $\lambda = 0^\circ$. Both observations disfavor the well-aligned model. The RVs and diffuser-assisted photometry are available as data behind the figure.}
\label{fig:diffuser}
\end{figure*}

To extract the RVs, we used a customized version of the Spectrum Radial Velocity Analyzer (SERVAL) pipeline \citep{zechmeister2018} optimized for NEID spectra, which uses the template-matching method to extract precise RVs. This version of the code is based both on the original SERVAL code by \cite{zechmeister2018} as well as the customized version for the Habitable-zone Planet Finder (HPF) instrument by \cite{stefansson2020}. We verified during NEID commissioning that when applied to the spectra of many reference stars, the SERVAL template-matching RVs are consistent with the official version of the NEID pipeline, which derives precise RVs using the cross-correlation method. However, the template-matching method is capable of using a higher fraction of the rich information content inherent in spectra of M dwarfs such as GJ\,3470, resulting in higher precision RVs.

As this was the first time we used this implementation of SERVAL for NEID data, we provide more details here. To extract precise RVs, we used order indices 30-104 spanning the wavelength region from 426\,nm to 895\,nm. Although NEID is sensitive down to 380\,nm, the bluer orders had very low SNR given the faintness of GJ\,3470 at blue wavelengths, and including those orders did not improve the resulting RV uncertainties. Barycentric corrections were calculated using the \texttt{barycorrpy} package \citep{kanodia2018}, which uses the algorithms from \cite{wright2014}. To mask out telluric lines, we used the synthetic telluric mask calculated using the TERRASPEC code \citep{bender2012}, an IDL wrapper to the Line-By-Line Radiative Transfer Model package \citep{clough2005}. We calculated the mask using parameters applicable for NEID's location on Earth, and nominal assumptions about humidity. We used this synthetic telluric spectrum to generate a thresholded binary mask. Any telluric line deeper than 0.5\% was masked. After generating the thresholded mask, we further broadened the mask by 21 wavelength resolution elements to conservatively mask telluric contaminated regions. We experimented with using the NEID sky fiber to subtract the background sky from the science fiber, but this did not significantly change the RVs. As we obtained a slightly higher RV precision without performing the sky-subtraction, we elected to extract the RVs from non-sky subtracted spectra.

\subsection{Diffuser-assisted Photometry}
To refine the transit ephemeris, we observed two photometric transits of GJ\,3470b using the Astrophysical Research Council Telescope Imaging Camera \citep[ARCTIC;][]{huehnerhoff2016} on the ARC 3.5m Telescope at Apache Point Observatory in New Mexico. The two nights were 2019 December 2 and 22 UT. We used the Engineered Diffuser available on the ARCTIC imager \citep{stefansson2017} because it enables high precision photometry by molding the point-spread function into a broad and stabilized top-hat shape. To minimize atmospheric systematics, we used a narrow-band (30\,nm wide) filter from Semrock Optics which is centered in a region with minimal telluric absorption at 857\,nm (for further details, see \citealt{stefansson2017} and \citealt{stefansson2018b}). The mean cadence was 19.1\,s and 18.2\,s for the two transits respectively. Both observations used ARCTIC's 2x2 binning mode, which has a gain of $2.0 \unit{e/ADU}$ and a plate scale of $0.22 \unit{\arcsec/pixel}$.

We bias corrected and flat fielded the data using standard procedures described by \cite{stefansson2017}. To extract the photometry, we used \texttt{AstroImageJ} \citep{collins2017}, following the methodology of \cite{stefansson2017}. We tried several different aperture sizes and ultimately adopted an aperture radius of 27 pixels and 34 pixels for the December 2 and 22 observations, respectively, which showed the lowest standard deviation in the transit residuals. For both observations, the background level was estimated from the counts within an annulus ranging from 60 to 90 pixels in radius. We estimated the uncertainty in each photometric measurement as the standard uncertainty from \texttt{AstroImageJ} accounting for photon, read, dark, and and digitization noise added in quadrature to our independent estimate of the scintillation noise following the methodology in \cite{stefansson2017}. The transits are shown in Figure \ref{fig:diffuser}.

\subsection{Out-of-transit Spectroscopy}
In addition to the in-transit spectroscopy with NEID, we analyzed out-of-transit RVs to constrain the possibility of an outer companion. For this analysis, we used RVs from the High Resolution Echelle Spectrometer \citep[HIRES;][]{vogt1994} on the Keck I telescope on Maunakea, the HARPS spectrograph \citep{mayor2003} on the 3.6m Telescope at La Silla Observatory in Chile, and the Habitable-zone Planet Finder (HPF) spectrograph \citep{mahadevan2012,mahadevan2014} on the 10m Hobby-Eberly Telescope in Texas.

For HIRES, we used precise RVs derived with the iodine technique as published by \cite{kosiarek2019}, totalling 56 RV points with a median RV precision of 1.89m/s. 

For HARPS, we note that \cite{kosiarek2019} analyzed RVs from HARPS from 2008 December to 2017 April including data from the original GJ 3470b discovery paper from \cite{bonfils2012}. In addition to the data analyzed in \cite{kosiarek2019}, we noticed that 6 additional RVs were publicly available on the HARPS archive\footnote{\url{http://archive.eso.org/eso/eso_archive_main.html}} obtained in 2018 March and April as part of program 198.C-0838(A) (PI: Bonfils). As these additional RVs extended the HARPS baseline by a year, we downloaded all of the available HARPS data from the HARPS archive, and extracted precise radial velocities using SERVAL. We removed one point as clear low S/N outlier, leaving 122 HARPS points with a median RV precision of 2.85m/s. We also used SERVAL to extract the H$\alpha$ activity index, and we extracted the Ca II H\&K $S_{\mathrm{HK}}$ index values from the HARPS spectra following \cite{gomesdasilva2011} and \cite{robertson2016}.

Additionally, we used near-infrared out-of-transit RVs obtained with the Habitable-zone Planet Finder (HPF) spectrograph. HPF is a fiber-fed near-infrared (NIR) spectrograph on the 10m Hobby-Eberly Telescope \citep{mahadevan2012,mahadevan2014} at McDonald Observatory in Texas, covering the $z$, $Y$, and $J$ bands from 810$\unit{nm}$-1260$\unit{nm}$ at a resolution of $R\sim55,000$. To enable precision radial velocities in the NIR, HPF is temperature stabilized at the milli-Kelvin level \citep{stefansson2016}. A subset of the HPF spectra were originally discussed by \cite{ninan2019gj3470} to demonstrate that GJ 3470b shows an absorption in the He 10830\AA\ line during transit. To avoid the complexity of modeling the RM effect, we only considered HPF data that were not obtained during transits. This resulted in 9 observations with a median RV precision of 4.5m/s. The HPF 1D spectra were reduced using the HPF pipeline following the procedures in \cite{ninan2018}, \cite{kaplan2018}, and \cite{metcalf2019}. Following the 1D spectral extraction, we reduced the HPF radial velocities using a version of the \texttt{SERVAL} template-matching RV-extraction pipeline \citep{zechmeister2018} optimized for HPF RV extractions, which is described in \cite{stefansson2020}. Following \cite{stefansson2020}, we only use the 8 HPF orders that are cleanest of tellurics, covering the wavelength regions from 8540-8890\AA, and 9940-10760\AA. We subtracted the estimated sky-background from the stellar spectrum using the dedicated HPF sky fiber, and we masked out telluric lines and sky-emission lines to minimize their impact on the RV determination.

Together the available RV data from HIRES, HARPS, and HPF span a baseline of 4709 days, or about 12.9 years from 2008 December 7 to 2021 October 29.

\section{Transit Ephemeris}
\label{sec:ephemeris}
To update the orbital ephemeris of GJ 3470b, we used a two-step procedure. First, we modeled the two ARCTIC transits independently to obtain precise transit midpoints for each transit. To model the transits, we followed the methodology of \cite{stefansson2020b} using the \texttt{juliet} code \citep{Espinoza2019}. \texttt{juliet} uses the \texttt{batman} code \citep{kreidberg2015} for the transit model and the \texttt{dynesty} dynamic nested sampler \citep{speagle2019} to perform a dynamic nested sampling of the posteriors. For the transit model, we placed broad uniform priors on the transit parameters $R_p/R_\star$, $a/R_\star$, and $b$. We sampled the limb darkening parameters using the quadratic $q_1$ and $q_2$ limb-darkening parameterization of \cite{kipping2013} with uniform priors. To obtain a constraint on the transit midpoint, we placed a Gaussian prior on the period of the planet based on the value reported by \cite{nascimbeni2013}, and a broad uniform prior on the transit midpoint. To account for correlated noise observed in the light curves, we used a Matern-3/2 Gaussian Process (GP) kernel implemented in \texttt{celerite} \citep{Foreman-Mackey2017} as available in the \texttt{juliet} code, placing broad uninformative priors on the GP hyperparameters. Figure \ref{fig:diffuser} shows the two transits along with the best-fit models. The transit midpoints (in the \unit{BJD_{TDB}} system) are $T_{C1} = 2458819.85859 \pm 0.00025$ and $T_{C2} = 2458839.87824 \pm 0.00045$.

Second, to update the transit ephemeris, we fitted a linear function of epoch number,
\begin{equation}
T_C  = n P + T_0,
\end{equation}
to the transit midpoints from ARCTIC and the transit midpoint from \cite{nascimbeni2013}. The results were $P=3.33665266 \pm 0.00000030 \unit{days}$, and $T_0 = 2456340.725588 \pm 0.00010 \unit{BJD_{TDB}}$. These observations led to an improvement in the precision of the period measurement by a factor of 5, translating to an uncertainty in the transit midpoint of only 0.5\,min for the nights of the two NEID spectroscopic observations.

\section{Out of Transit RVs}
\label{sec:rv}

\begin{figure*}[t!]
\begin{center}
\includegraphics[width=0.7\textwidth]{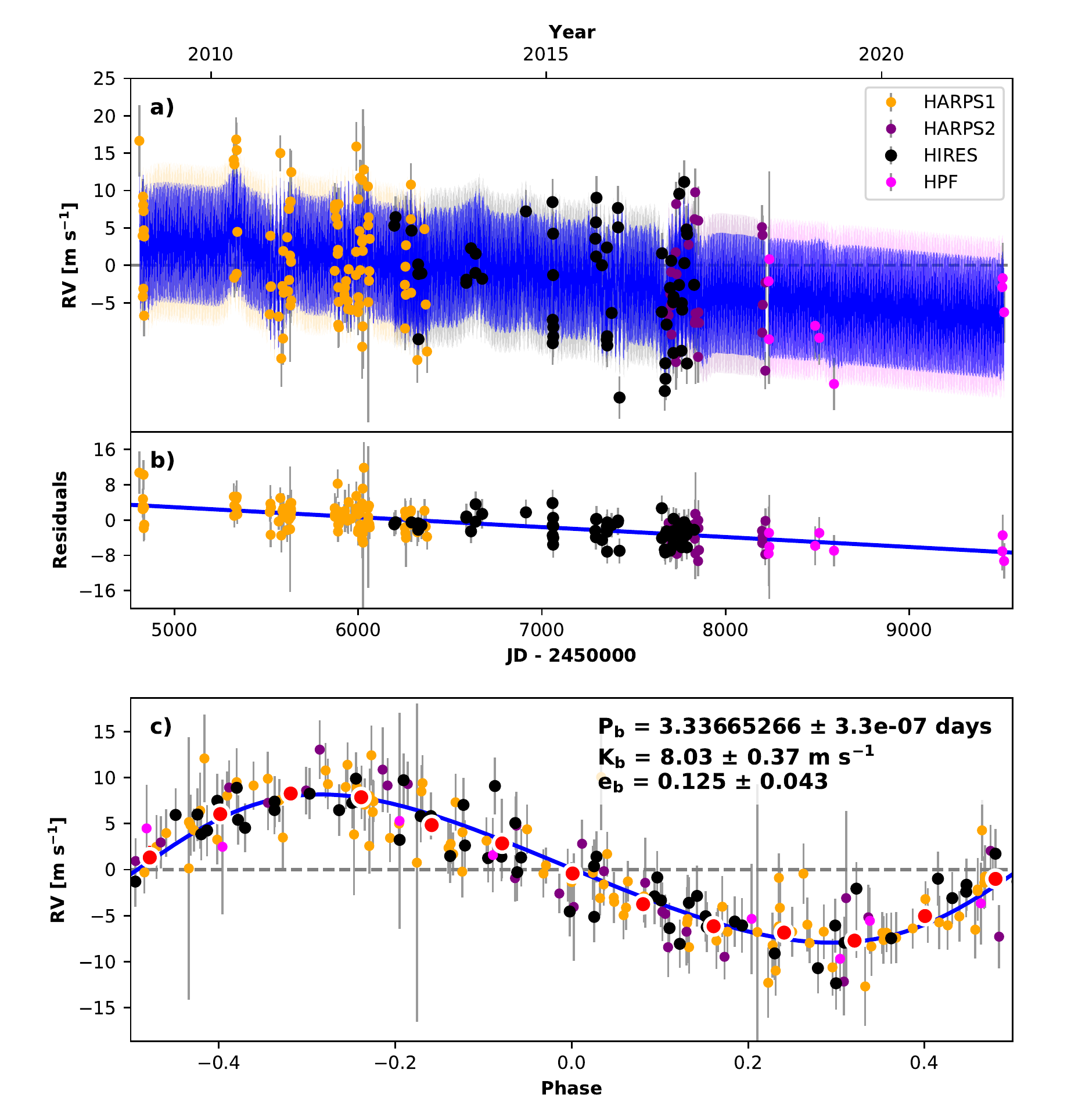}
\vspace{-0.5cm}
\end{center}
\caption{Fit of out-of-transit RVs of GJ 3470 with the known orbit of GJ 3470b, a linear slope, and a quasi-periodic GP model. HARPS RVs before and after the 2015 fiber break are shown in yellow (denoted HARPS1) and purple (denoted HARPS2), respectively, HIRES RVs are shown in black, and HPF RVs in pink. a) RVs as a function of time. b) Residuals including the RV slope. c) Phase-folded RVs. The RVs are available as machine readable data behind the figure.}
\label{fig:rv}
\end{figure*}

\subsection{RV Fit}
To precisely constrain the parameters of GJ 3470b and to probe for any evidence of an outer long-term companion, we modeled the available out-of-transit RVs of GJ\,3470 from HARPS, HIRES, and HPF. We fit the RVs using the \texttt{radvel} code \citep{fulton2018}. 

\begin{deluxetable*}{lccccc}
\tablecaption{Summary of out-of-transit RV fits. \label{tab:rvbic}}
\tablehead{\colhead{Fit}         & \colhead{Slope (m/s/day)}              &  \colhead{BIC}        & \colhead{$\Delta$BIC}  & \colhead{AIC}       & \colhead{$\Delta$AIC}}
\startdata 
Model I: No Slope, no GP         & -                                      &  1129.9               & 0.0                    & 1090.0              & 0.0     \\
Model II: Slope, no GP           & $-0.0023\pm0.00069$                    &  1123.7               & -6.2                   & 1080.9              & -9.1    \\
Model III: No Slope, GP          & -                                      &  1124.0               & -6.0                   & 1072.6              & -17.4   \\
Model IV: Slope, GP              & $-0.0022\pm0.0011$                     &  1124.6               & -5.4                   & 1070.5              & -19.6   \\
\enddata
\end{deluxetable*}

To assess the statistical significance of a possible RV slope and quantify the need for using a Gaussian Process (GP) red-noise model to account for stellar activity signatures in the overall RV dataset, we performed a total of four fits with and without RV slopes and with and without a GP model:
\begin{itemize}
\item \textbf{Model I:} No Slope, no GP,
\item \textbf{Model II:} Slope, no GP,
\item \textbf{Model III:} No Slope, GP,
\item \textbf{Model IV:} Slope, GP.
\end{itemize}
These models are summarized in Table \ref{tab:rvbic}. For all fits, we placed Gaussian priors on the orbital period ($P$) and transit midpoint ($T_c$) from our derived ephemeris in Section \ref{sec:ephemeris}, a Gaussian prior on the $e\cos \omega = 0.014546_{-0.000659}^{+0.000753}$ from \cite{kosiarek2019} which was derived from the timing of Spitzer secondary eclipse measurements presented by \cite{benneke2019}. We placed uniform priors on the RV semi-amplitude ($K$). The HARPS fiber link was upgraded from circular to octagonal fibers on 2015~May~28 which introduced a zero-point offset in the RV timeseries \citep{locurto2015}. To account for this offset, we modeled the HARPS RVs as two separate RV streams. We include independent RV offset and instrument jitter parameters for each dataset (HARPS before upgrade, HARPS after upgrade, HIRES and HPF). Following the procedures in the \texttt{radvel} package, we first obtained the global maximum likelihood solution, and then sampled the posteriors surrounding the maximum likelihood solution using the \texttt{emcee} affine-invariant Markov Chain Monte Carlo (MCMC) sampler \citep{dfm2013}. We used the default MCMC convergence criteria within the \texttt{radvel} package: convergence is reached when the Gelman-Rubin statistic is less than 1.01 and that the number of independent samples is greater than 1000 for all free parameters.

\begin{deluxetable*}{llcc}
\tablecaption{Summary of priors and resulting posteriors for the out-of-transit RV analysis. $\mathcal{N}(m,\sigma)$ denotes a normal prior with mean $m$, and standard deviation $\sigma$; $\mathcal{J}(a,b)$ denotes a Jeffreys prior with a start value $a$ and end value $b$. \label{tab:rvparams}.}
\tablehead{\colhead{~~~Parameter}&                             \colhead{Description}  &      \colhead{Prior}                      &    \colhead{Posterior}}
\startdata
\multicolumn{4}{l}{\hspace{-0.3cm} MCMC Input Parameters:}           \\
 $T_{C}$ $(\mathrm{BJD_{TDB}})$ &                                    Transit midpoint &  $\mathcal{N}(2456340.72559,0.00010)$$^a$ &  $2456340.72559_{-0.0001}^{+0.00011}$    \\ %
                            $P$ &                               Orbital period (days) &  $\mathcal{N}(3.33665267,0.00000032)$$^a$ &  $3.33665266_{-0.0000003}^{+0.0000003}$  \\ %
               $e \cos(\omega)$ &            Eccentricity and Argument of periastron  &      $\mathcal{N}(0.014546,0.0007)$$^b$   &  $0.01444_{-0.00074}^{+0.00074}$         \\ %
               $e \sin(\omega)$ &            Eccentricity and Argument of periastron  &      -                                    &  $-0.125_{-0.043}^{+0.043}$              \\ %
                            $K$ &                            RV semi-amplitude (m/s)  &      -                                    &  $8.03_{-0.37}^{+0.38}$                  \\ %
                 $\dot{\gamma}$ &                                 RV slope (m/s/day)  &      -                                    &  $-0.0022_{-0.0011}^{+0.0011}$           \\ %
      $\gamma_{\mathrm{HIRES}}$ &                               HIRES RV offset (m/s) &      -                                    &  $2.0_{-1.3}^{+1.3}$                     \\ %
     $\gamma_{\mathrm{HARPS1}}$ &          HARPS RV offset before fiber upgrade (m/s) &      -                                    &  $-0.11_{-0.89}^{+0.83}$                 \\ %
     $\gamma_{\mathrm{HARPS2}}$ &           HARPS RV offset after fiber upgrade (m/s) &      -                                    &  $-1.0_{-2.1}^{+2.1}$                    \\ %
        $\gamma_{\mathrm{HPF}}$ &                                HPF RV offset  (m/s) &      -                                    &  $6.5_{-3.4}^{+3.6}$                     \\ %
      $\sigma_{\mathrm{HIRES}}$ &                               HIRES RV Jitter (m/s) &      $\mathcal{J}(0.01,10)$               &  $2.07_{-0.76}^{+0.67}$                  \\ %
     $\sigma_{\mathrm{HARPS1}}$ &          HARPS RV Jitter before fiber upgrade (m/s) &      $\mathcal{J}(0.01,10)$               &  $0.027_{-0.014}^{+0.056}$               \\ %
     $\sigma_{\mathrm{HARPS2}}$ &           HARPS RV Jitter after fiber upgrade (m/s) &      $\mathcal{J}(0.01,10)$               &  $0.03_{-0.016}^{+0.073}$                \\ %
        $\sigma_{\mathrm{HPF}}$ &                                 HPF RV Jitter (m/s) &      $\mathcal{J}(0.01,10)$               &  $0.028_{-0.014}^{+0.060}$               \\ %
                       $\eta_1$ &                                  GP amplitude (m/s) &      $\mathcal{J}(0.01,100)$              &  $2.92_{-0.43}^{+0.47}$                  \\ %
                       $\eta_2$ &                               GP periodicity (days) &      $\mathcal{N}(21.54,0.49)$$^c$        &  $21.64_{-0.43}^{+0.45}$                 \\ %
                       $\eta_3$ &                        GP exponential length scale  &      $\mathcal{N}(49,8)$$^c$              &  $50.1_{-8.0}^{+8.1}$                    \\ %
                       $\eta_4$ &                  GP periodicity length scale (days) &      $\mathcal{N}(0.55,0.06)$$^c$         &  $0.548_{-0.057}^{+0.058}$               \\ %
\multicolumn{4}{l}{\hspace{-0.3cm} Derived Parameters:}  \\
                            $e$ &                                        Eccentricity &      -                                    &  $0.125_{-0.042}^{+0.043}$               \\ %
                       $\omega$ &                        Argument of Periastron (deg) &      -                                    &  $-83.4_{-1.7}^{+3.4}$                   \\ %
                    $m_b\sin i$ &                       Mass of GJ 3470b ($M_\oplus$) &      -                                    &  $12.14_{-0.66}^{+0.68}$                 \\ %
\enddata
\vspace{-0.1cm}
\tablenotetext{a}{Priors on ephemeris is from Section \ref{sec:ephemeris}.}
\vspace{-0.2cm}
\tablenotetext{b}{Prior on $e\cos \omega$ is from \cite{kosiarek2019}.}
\vspace{-0.2cm}
\tablenotetext{c}{Priors on GP parameters are from \cite{kosiarek2019}.}
\end{deluxetable*}

For the fits with an RV slope parameter, we placed no prior on the slope. For fits using a GP, we used the quasi-periodic GP kernel as implemented in the \texttt{george} package \citep{ambikasaran2015} available in \texttt{radvel}. This GP kernel has four hyper parameters: the GP RV amplitude $\eta_1$, a periodicity parameter $\eta_2$, length scale of the exponential component $\eta_3$, and a length scale of the periodic component $\eta_4$. We follow \cite{kosiarek2019}, and we placed Gaussian priors on $\eta_2=49.0 \pm 8.0$, $\eta_3=21.54 \pm 0.49 \unit{days}$, and $\eta_4 = 0.55 \pm 0.06$, using the values from \cite{kosiarek2019} which they constrained by modeling photometric data from Fairborn Observatory. We placed a uniform prior on the GP amplitude.

Table \ref{tab:rvbic} compares the Bayesian Information Criterion (BIC), and the Aikake Information Criterion (AIC) for the four fits we considered. Both the BIC and AIC measure model likelihood while penalizing a higher number of free parameters, where the AIC is less punative toward the number of free parameters. From Table \ref{tab:rvbic}, we see that Models II, III and IV are significantly favored over Model I with $\Delta\mathrm{BIC}\geq5.4$ and $\Delta\mathrm{AIC}\geq9.1$ in favor of Models II, III and IV. We see that the two models that have a slope (Model II and IV) yield consistent slope values of $\dot{\gamma} = -0.0023 \pm 0.00069 \unit{m/s/day}$ and $\dot{\gamma} = -0.0022 \pm 0.0011 \unit{m/s/day}$. From Table \ref{tab:rvbic} we see that Models II, III and IV all have similar BIC values within $\Delta\mathrm{BIC}\sim0.9$ of each other, suggesting they are statistically indistinguishable. The simplest of these models, Model II with a Keplerian and a simple slope, is a good description of the data. We further note that the AIC---which penalizes for additional fitting parameters to a lesser extent than the BIC---favors models with the GP included (Models III and IV). As there are independent evidence of stellar activity from the RVs directly (see discussion in \citealt{kosiarek2019} and in Section \ref{sec:activity}), fits that include a GP stellar activity model are warranted. As the AIC is the lowest for Model IV, which explicitly accounts for a long-term RV slope and signatures of stellar activity, we formally adopt those values, which suggest the data are compatible with an RV slope with $\dot{\gamma} = -0.0022 \pm 0.0011 \unit{m/s/day}$, suggesting a detection of a long-term RV slope at $2\sigma$ confidence. We urge additional RV follow-up to confirm or refute this candidate RV slope.

Figure \ref{fig:rv} shows the resulting RV fit from Model IV, along with the phase-folded RVs, and Table \ref{tab:rvparams} summarizes the priors and the resulting posteriors. We obtain a semi-amplitude of \resK\ which is consistent with the semi-amplitude of $K=8.21_{-0.46}^{+0.47} \unit{m/s}$ reported by \cite{kosiarek2019}.

\begin{figure*}[t!]
\begin{center}
\includegraphics[width=0.9\textwidth]{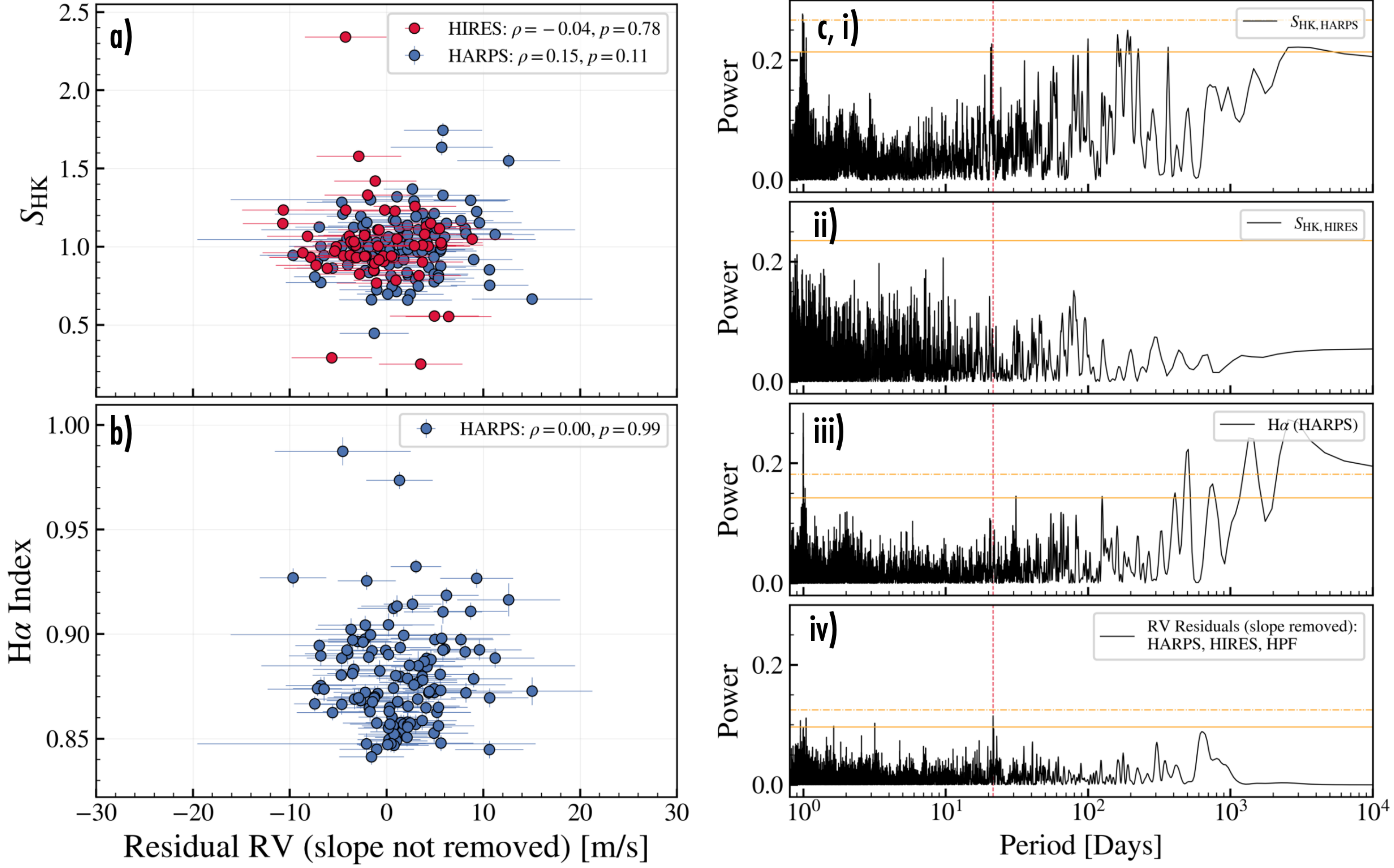}
\vspace{-0.4cm}
\end{center}
\caption{a) $S_{\mathrm{HK}}$ as a function of the residual radial velocity from Model II. The residual RV includes the RV slope (see Figure \ref{fig:rv}b). b) H$\alpha$ index as a function of residual RV. The Spearman's rank correlation coefficient $\rho$ for each dataset is shown in the legends. We see no clear correlations with $S_{\mathrm{HK}}$ or H$\alpha$ with the residual RVs. c) Lomb-Scargle periodograms of i) the $S_{\mathrm{HK}}$ indicator from HARPS, ii) $S_{\mathrm{HK}}$ indicator from HIRES, iii) H$\alpha$ indicator from HARPS, iv) residual RV (slope removed) from HARPS, HIRES and HPF. The known stellar rotation period is shown with the red dashed line. The orange horizontal lines show the 10\% (solid line) and the 1\% (dot-dashed line) false alarm probabilities calculated with the bootstrap method.}
\label{fig:activityrv}
\end{figure*}

\subsection{Stellar Activity Correlations}
\label{sec:activity}
To investigate if stellar activity could account for the long-term RV slope, we studied both the Mount Wilson $S_{\mathrm{HK}}$ index, which traces the chromospheric emission in the cores of the Ca II H\&K lines \citep{vaughan1978}, and the H$\alpha$ index as calculated by the SERVAL pipeline which probes the emission of the H$\alpha$ line. Figure \ref{fig:activityrv}a and b) show the $S_{\mathrm{HK}}$ and H$\alpha$ indices as a function of the residual radial velocities from Model II before taking out the long-term RV slope. We show the residual RVs from Model II, as that model does not use a GP that could potentially remove long-term stellar activity activity effects seen in the RVs. The Spearman's rank correlation coefficient between the residual RVs and the $S_{\mathrm{HK}}$ values is $\rho_{\mathrm{HIRES}}=-0.04$ and $\rho_{\mathrm{HARPS}}=0.15$ with $p$ values $p=0.78$ and $p=0.11$ for the HIRES and HARPS data, respectively. The Spearman's rank correlation coefficient between the H$\alpha$ index and the HARPS residual RVs is $\rho = 0.0$ with $p=0.99$. We see that all cases suggest there is no correlation between the activity indices and the the residual RVs including the RV slope, disfavoring a stellar-activity origin for the long-term RV slope.

To investigate activity signatures at shorter timescales, in particular at the known stellar rotation period, Figure \ref{fig:activityrv} shows Generalized Lomb-Scargle periodograms of the $S_{\mathrm{HK}}$ and H$\alpha$ activity indicators as well as the residual RVs after removing the RV slope from Model II. We see peaks close to the known stellar rotation period in the $S_{\mathrm{HK}}$ index in HARPS (Figure \ref{fig:activityrv}c-i), and in the RV residuals (Figure \ref{fig:activityrv}c-iv) with false alarm probabilities $\leq10\%$. Similar peaks in the RV residuals at the stellar rotation period were noted by \cite{kosiarek2019}, suggesting that they originate from stellar active regions. We conclude similar to \cite{kosiarek2019} that this further motivates the use of a GP model to account for this impact in the RVs. From the parameters from Model IV in Table \ref{tab:rvparams}, we see that the best-fit GP amplitude is $3.0 \pm 0.5 \unit{m/s}$.

\begin{deluxetable*}{llccc}
\tablecaption{Summary of priors and resulting posteriors for the RM analysis. $\mathcal{N}(m,\sigma)$ denotes a normal prior with mean $m$, and standard deviation $\sigma$; $\mathcal{U}(a,b)$ denotes a uniform prior with a start value $a$ and end value $b$.\label{tab:planetparams}. }
\tablehead{\colhead{~~~Parameter}&                             \colhead{Description}  &      \colhead{Prior}                      &    \colhead{Posterior}                    &    \colhead{Notes}. }
\startdata
\multicolumn{5}{l}{\hspace{-0.3cm} MCMC Input Parameters:}           \\
 $T_{C}$ $(\mathrm{BJD_{TDB}})$ &                                    Transit midpoint &      $\mathcal{N}(2456340.72559,0.00010)$ &  $2456340.725592_{-0.00010}^{+0.00010}$   & This work \\
                            $P$ &                               Orbital period (days) &      $\mathcal{N}(3.33665267,0.00000032)$ &  $3.33665267_{-0.0000003}^{+0.0000003}$   & This work \\
                      $R_p/R_*$ &                                       Radius ratio  &      $\mathcal{N}(0.07642,0.00037)$       &  $0.07642_{-0.00036}^{+0.00037}$          & \cite{biddle2014}    \\
                        $a/R_*$ &                             Scaled semi-major axis  &      $\mathcal{N}(13.94,0.5)$             &  $13.99_{-0.47}^{+0.47}$                  & \cite{biddle2014}    \\
                            $i$ &                     Transit inclination ($^\circ$)  &      $\mathcal{N}(88.88,0.5)$             &  $88.86_{-0.42}^{+0.38}$                  & \cite{biddle2014}    \\
                            $e$ &                                       Eccentricity  &      $\mathcal{N}(0.125,0.042)$           &  $0.123_{-0.040}^{+0.040}$                & This work \\
                       $\omega$ &                  Argument of periastron ($^\circ$)  &      $\mathcal{N}(-83.4,3)$               &  $-83_{-3}^{+3}$                          & This work \\
                            $K$ &                            RV semi-amplitude (m/s)  &      $\mathcal{N}(8.0,0.37)$              &  $7.98_{-0.36}^{+0.37}$                   & This work \\
                   $\gamma_{1}$ &                      NEID RV offset Transit 1 (m/s) &      $\mathcal{U}(-100,100)$              &  $0.5_{-0.45}^{+0.44}$                    & This work \\
                   $\gamma_{2}$ &                      NEID RV offset Transit 2 (m/s) &      $\mathcal{U}(-100,100)$              &  $-0.39_{-0.78}^{+0.78}$                  & This work \\
                          $u_1$ &                     Linear limb darkening parameter &      $\mathcal{N}(0.35,0.1)$              &  $0.347_{-0.10}^{+0.099}$                 & This work \\
                          $u_2$ &                  Quadratic limb darkening parameter &      $\mathcal{N}(0.32,0.05)$             &  $0.321_{-0.051}^{+0.050}$                & This work \\
                        $\beta$ &                Intrinsic stellar line width (km/s)  &      $\mathcal{N}(3.0,0.5)$               &  $3.0_{-0.5}^{+0.5}$                      & This work \\
                      $\lambda$ &                      Sky-projected obliquity (deg)  &      $\mathcal{U}(-180,180)$              &  \reslambdaval                            & This work \\ 
                      $R_\star$ &                                     Radius of star  &      $\mathcal{N}(0.5,0.016)$             &  $0.499_{-0.016}^{+0.016}$                & This work \\ 
                      $P_{rot}$ &                    Stellar rotation period (days)   &      $\mathcal{N}(21.54,0.49)$            &  $21.56_{-0.49}^{+0.49}$                  & This work \\ 
                      $\cos i$  &                      Cosine of Stellar inclination  &      $\mathcal{U}(0,1)$                   &  $0.69_{-0.40}^{+0.21}$                   & This work \\
\multicolumn{5}{l}{\hspace{-0.3cm} Derived Parameters:}  \\
                     $v \sin i$ &               Projected Rotational Velocity (km/s)  &      -                                    &  \resvsinival                             & This work \\ 
                     $i_\star$  &                      Stellar inclination (deg)      &      -                                    &  \resistarval                             & This work \\
                         $\psi$ &                                True Obliquity (deg) &      -                                    &  \respsival                               & This work \\
\enddata
\end{deluxetable*}

\section{RM Effect}
\label{sec:rm}
To model the RM effect observations, we broadly followed the methodology of \cite{stefansson2020b}, which we have implemented in a code named \texttt{rmfit}. In short, we use the RM effect model framework from \cite{hirano2011} along with the \texttt{radvel} code \citep{fulton2018} to account for the orbital motion of the planet during the transit. We jointly modeled both NEID transits. We placed Gaussian priors on the planet parameters which have been precisely constrained from other observations, and we placed informative priors on the ephemeris derived in Section \ref{sec:ephemeris}. We place broad uniform priors on $\lambda$. To constrain the true obliquity, we additionally sample the stellar inclination ($i_\star$), the stellar radius ($R_\star$), and the stellar rotation period ($P_{\mathrm{rot}}$). We place Gaussian priors on the known stellar radius and rotation period to calculate the equatorial velocity of the star, $v_{\mathrm{eq}}$, and we sample the stellar inclination sampled as $\cos i_\star$ with a uniform prior on $\cos i_\star$. We then estimate the $v \sin i = v_{\mathrm{eq}} \sin i_\star = v_{\mathrm{eq}}\sqrt{1-\cos^2 i_\star}$. This broadly follows the methodology in \cite{masuda2020}, to account for the fact that $v\sin i$ and $v_{\mathrm{eq}}$ are not independent variables (e.g., $v \sin i$ is always lower than $v_{\mathrm{eq}}$). To calculate the true obliquity $\psi$, we used the geometric relation,
\begin{equation}
\cos \psi = \sin i_\star \cos \lambda \sin i + \cos i_\star \cos i,
\label{eq6:psi}
\end{equation}
where $i_\star$ is the stellar inclination, $i$ is the orbital inclination, and $\lambda$ is the sky-projected obliquity.

To account for any possible systematics on timescales longer than one night, we allowed for a separate RV offset for each transit. We placed informative priors on the limb-darkening parameters corresponding to the expected range for the $R$ and $I$-band\footnote{We estimated the limb-darkening parameters using the EXOFAST web-applet: \url{https://astroutils.astronomy.osu.edu/exofast/limbdark.shtml}}, where the bulk of the RV information content is located for these observations. We assumed a quadratic limb-darkening law. To account for the finite exposure times of our RV observations, we super-sampled the model with 86-second sampling (7-fold sampling) and averaged the model into 600-seconds bins before comparing the model to the data. We set the intrinsic line width $\beta$ to the width of the NEID resolution element, i.e., $\beta = 3.0 \pm 0.5 \unit{km/s}$, where the uncertainty is meant to account for any effects of macroturbulence or other processes that could broaden the line profile. For the RM fit, we ignore any effects of stellar activity due to the short duration of the transit compared to the stellar rotation period.

Before MCMC sampling, we first obtained a global maximum-likelihood solution using the using the \texttt{PyDE} differential evolution optimizer \citep{pyde}. We then initialized 60 MCMC walkers in the vicinity of the global most probable solution using the \texttt{emcee} MCMC affine-invariant sampling package \citep{dfm2013}. We ran the 60 walkers for 50{,}000 steps. The mean integrated correlation time for the parameters was 330, suggesting the 50,000 MCMC steps should be sufficiently sampling the posterior distribution. Further, after removing the first 2{,}000 burn-in steps, the Gelman-Rubin statistic of the resulting chains was within $1\%$ of unity, which we consider well-mixed. Table \ref{tab:planetparams} summarizes the priors and resulting posteriors.

\begin{figure*}[t!]
\begin{center}
\includegraphics[width=0.8\textwidth]{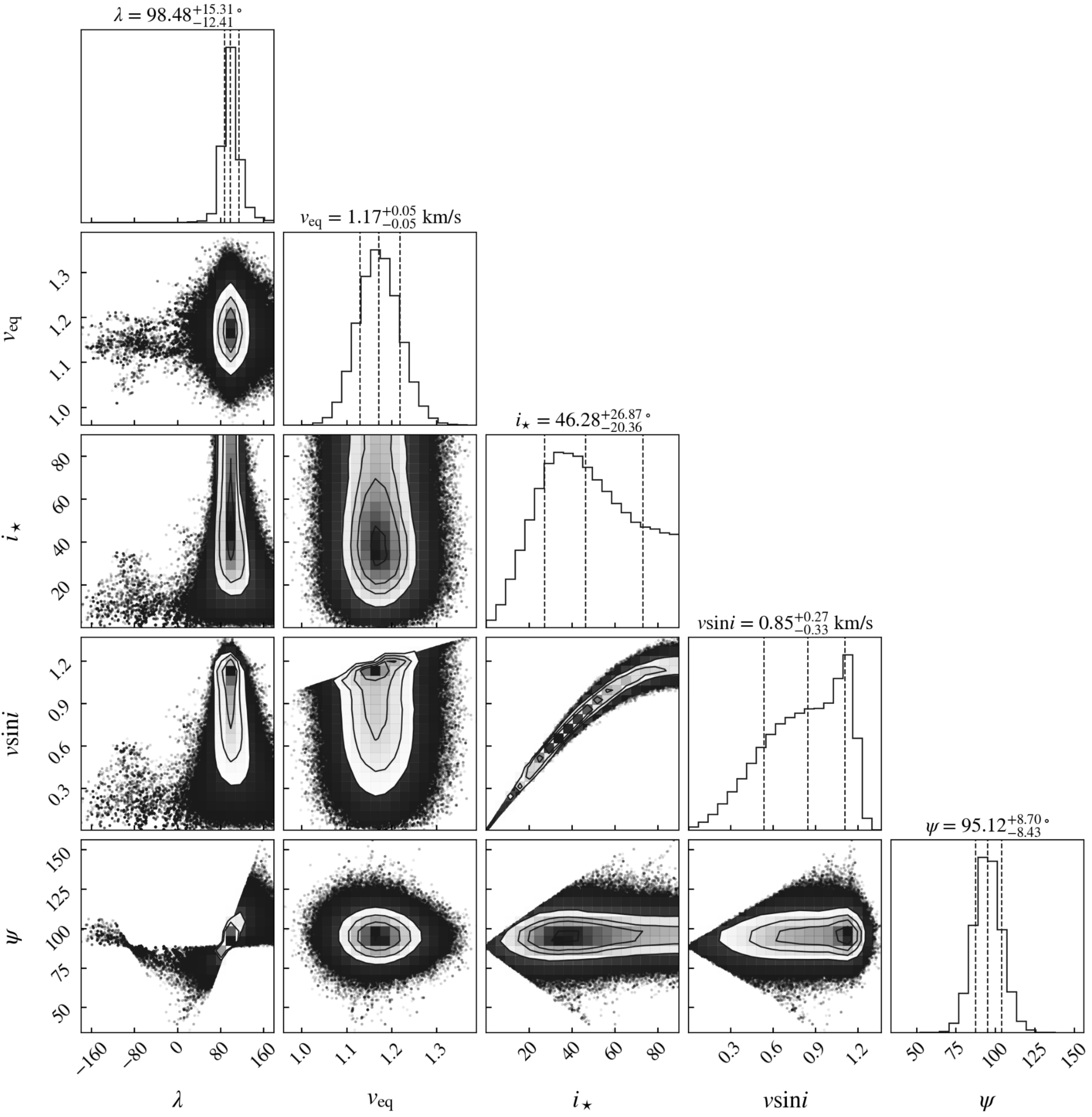}
\vspace{-0.5cm}
\end{center}
\caption{Corner plot of the posteriors of $\lambda$, $v_{\mathrm{eq}}$, $i_\star$, $v \sin i$, and $\psi$. The posteriors for the true obliquity are \respsi, and are compatible with $90^\circ$ for any posterior value of $v \sin i$.}
\label{fig:corner}
\end{figure*}

Figure \ref{fig:diffuser}b and c shows the RM effect observations along with the best-fit model (red) which yields a sky-projected obliquity of \reslambda, \resvsini, stellar inclination of \resistar, and a true obliquity of \respsi. Figure \ref{fig:corner} shows a corner plot of the posteriors, which shows that for any values of $v \sin i$, the true obliquity is robustly $\sim$90$^\circ$, suggesting a polar orbit. From the posteriors, it is valuable to examine Figure \ref{fig:corner} in the limit of low and high values of $v \sin i$:
\begin{itemize}
\item High values of $v \sin i$: In this case, $v \sin i \sim v_{\mathrm{eq}}$, which represents the highest posterior probability solution as we see from Figure \ref{fig:corner}. In this limit, we see that for the highest values of $v \sin i$, both $\lambda$ and $\psi$ are confidently $\sim$90$^\circ$. As an additional comparison, in Figure \ref{fig:diffuser}, we compare the best-fit RM model in red to the expected well-aligned model in gray which assumes $\psi=0^\circ$ (i.e., where we fix $\lambda=0^\circ$ and $v\sin i = v_{\mathrm{eq}}$). The Bayesian Information Criterion (BIC) for the best-fit model is $\mathrm{BIC}=44.5$ with 44 degrees of freedom. The BIC for the well-aligned model is $\mathrm{BIC}=103.1$ with 46 degrees of freedom. The resulting $\Delta$BIC = 58.6 strongly disfavors the well-aligned model relative to the best-fit model.
\item Low values of $v \sin i$: From Figure \ref{fig:corner} we see that although the posterior probability of $v \sin i$ vanishes at $0 \unit{km/s}$, $v \sin i$ is still compatible with low $v \sin i$ values of a few hundred m/s. For such low $v\sin i$ values, we see that the constraint on $\lambda$ becomes poorer. However, as $v \sin i$ decreases, $i_\star$ has to decrease accordingly which in turn maintains $\psi\sim90^\circ$. The posteriors show that in the limit of the lowest $v \sin i$ values, $\psi$ becomes even more tightly constrained to $90^\circ$ than at higher values of $v \sin i$.
\end{itemize}

\begin{figure*}[t!]
\begin{center}
\includegraphics[width=1.0\textwidth]{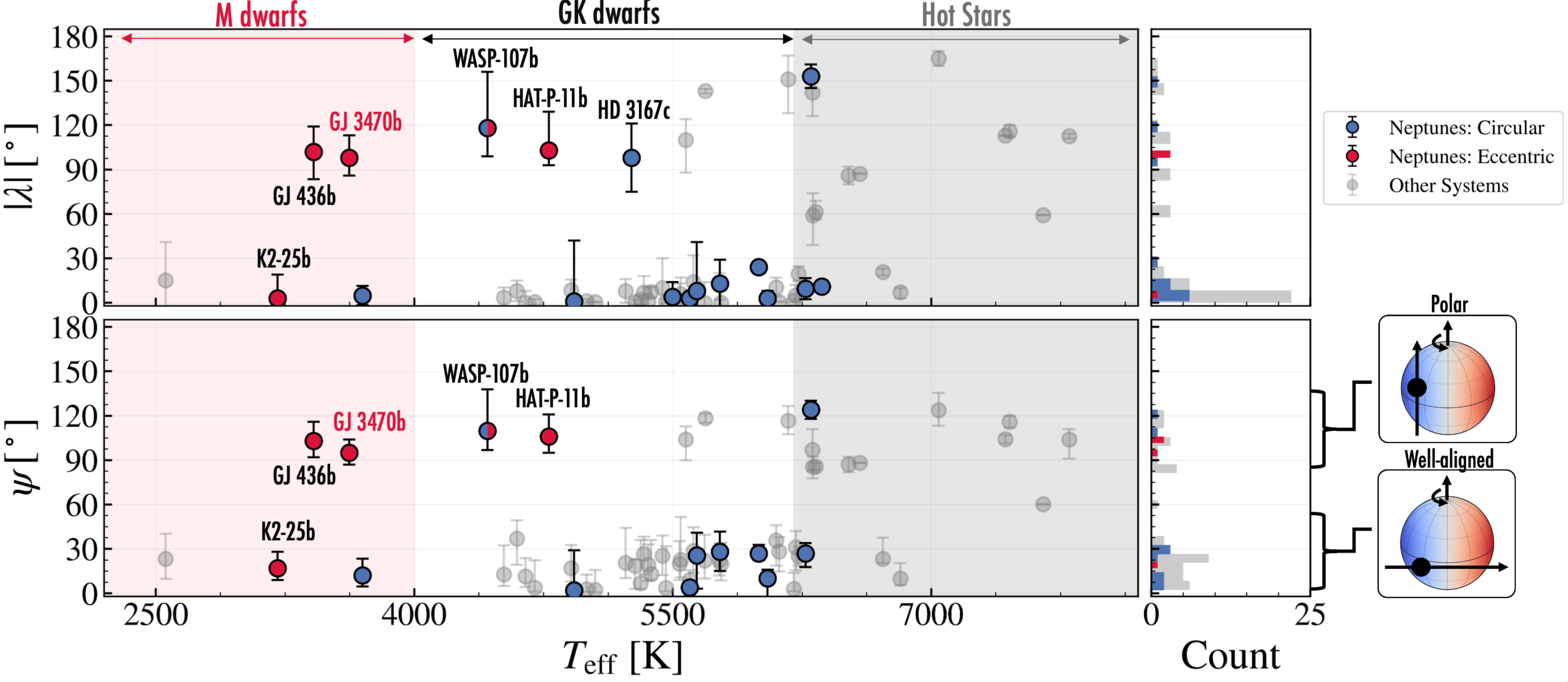}
\vspace{-0.9cm}
\end{center}
\caption{\textbf{Polar Neptunes:} Sky-projected obliquities ($\lambda$; upper panel) and true obliquities ($\psi$; lower panel; when available) as a function of stellar effective temperature. Planets either similar in radius or mass to Neptune ($1.5 R_\oplus < R < 6 R_\oplus$ or $10 M_\oplus < M < 50 M_\oplus$) on circular orbits are highlighted in blue, and Neptunes on eccentric orbits ($e>0.1$ and >2$\sigma$ discrepant from $e=0$) are shown in red. Other systems from \cite{albrecht2021} are shown in grey. We show WASP-107b as both red and blue, as it formally has an eccentricity constraint of $e=0.06\pm0.04$ and \cite{piaulet2021} mention it could have a moderate eccentricity. GJ\,3470b joins a growing sample of warm Neptunes orbiting cool stars on polar orbits. Data obtained from \cite{albrecht2021}, TEPCAT \citep{southworth2011}, and the NASA Exoplanet archive \citep{akeson2013}.}
\label{fig:rm}
\end{figure*}

To test the robustness of the results, we performed three additional tests. First, we tried placing a uniform prior on the semi-amplitude $K$ of the planet from 0 to 25m/s instead of an informative prior. This resulted in fully consistent results to those presented in Table \ref{tab:planetparams}. Second, we performed a separate RM fit where we only fit the parameters that are primarily constrained by the RM curve ($\lambda$, $v\sin i$, and the two RV offsets), while keeping the other values fixed to their median best-fit values or most likely prior values. This also resulted in fully consistent parameters with those presented in Table \ref{tab:planetparams}. Third, we also tried a fit with a uniform prior on the $v \sin i$, which yielded fully consistent parameters. Lastly, we experimented fitting the two transits separately, which yielded fully consistent parameters although with lower significance. We adopt the posteriors from the joint fit, as that fit leverages information from both transit observations.

\section{Discussion}
\label{sec:discussion}

\subsection{Neptunes in Eccentric Polar Orbits around Cool Stars}
Recently, \cite{albrecht2021} noticed that the planets with projected obliquities larger than about $40^\circ$ show an apparent preference for polar orbits ($\psi = 80-125^\circ)$ rather than spanning the full range of possible obliquities. Most of the polar systems involve hot Jupiters ($R > 6 R_\oplus$ and $P<10$~days) because the RM measurements of smaller and longer-period planets are more difficult.

Among the sample from \cite{albrecht2021} is a collection of four warm Neptunes ($a/R_\star \gtrsim 8$; $1.5 < R_p/R_\oplus < 6$) orbiting cool K and M dwarfs that are observed to be on polar orbits: HAT-P-11b \citep{winn2010,hirano2011hatp11}, GJ\,436b \citep{bourrier2018,bourrier2022}, HD 3167c \citep{dalal2019,bourrier2021}, and WASP-107b \citep{DaiWinn2017,rubenzahl2021}. Figure \ref{fig:rm} highlights these planets along with obliquity constraints available for other Neptunes\footnote{Obliquity constraints retrieved from the TEPCAT database \citep{southworth2011} in August 2021.}. Together with GJ 3470b, we have a sample of five polar warm Neptunes. Four of these planets---GJ 3470b, GJ 436b, HAT-P-11b, and WASP-107b---all reside in or at the edge of the ``Neptune Desert'' \citep[e.g.,][]{mazeh2016,owen2018} and are observed to have evaporating atmospheres. Further, three of these planets have non-circular orbits with eccentricities above 0.1 (GJ 3470b, GJ 436b, and HAT-P-11b), and we note that WASP-107 has an eccentricity constraint of $e=0.06\pm0.04$ and \cite{piaulet2021} explicitly mention the planet could have a moderate eccentricity. These similar properties could suggest that these systems share a similar formation history. We discuss possible formation scenarios in Section \ref{sec:formation}.

\subsection{Tidal Inflation}
It is intriguing to consider possible causal links between the orbital properties of the polar, moderately eccentric Neptunes mentioned above and their atmospheric mass loss. One possibility is that their mass loss is enhanced from atmospheric inflation driven by tidal heating, with the atmosphere being eroded by photoevaporation \citep[see e.g.,][]{owen2018,attia2021}. Short-period planets experience significant tidal deformations due to their close proximity to their host stars. Whenever the planets have non-zero eccentricities and/or axial tilts (planetary obliquities), these tidal deformations are non-uniform along the orbital path, and they drive interior friction and heat dissipation known as ``tidal heating'' \citep[e.g.,][]{Jackson2008}. This extra interior energy source can significantly inflate the atmospheres of gas-rich planets, as shown in previous works on hot Jupiters \citep[e.g.,][]{Bodenheimer2001, Miller2009}, sub-Neptunes \citep{Millholland2019}, and sub-Saturns \citep{Millholland2020}.

\begin{figure}[t!]
\begin{center}
\includegraphics[width=1.0\columnwidth]{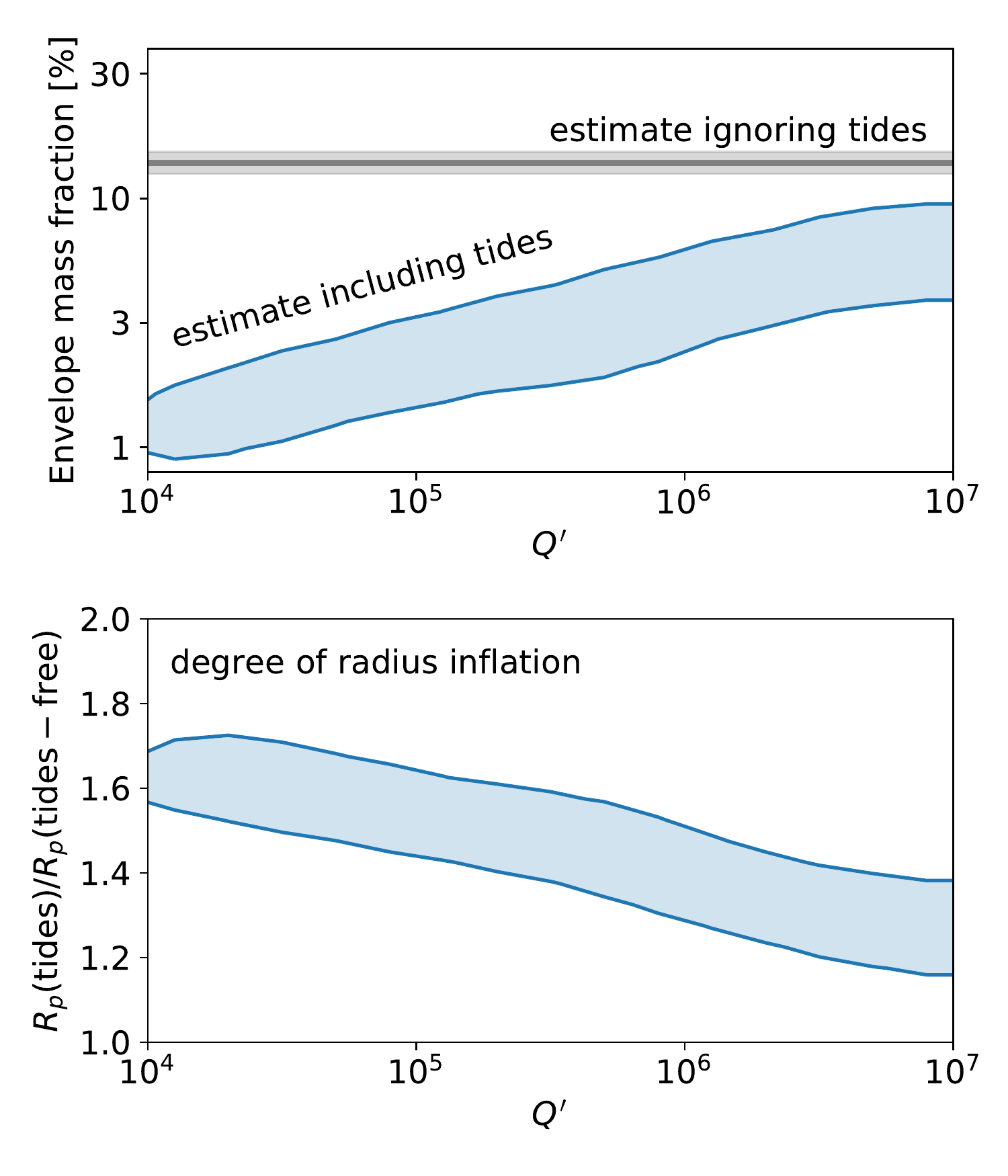}
\vspace{-0.9cm}
\end{center}
\caption{Results from the tidal model fit to the observed properties of GJ 3470b. Top panel: Comparison of the envelope mass fraction ($f_{\mathrm{env}}\equiv M_{\mathrm{env}}/M_p$) when tides are included and when they are ignored. The gray solid line and banding indicates the mean and standard deviation of the tides-free estimate, $f_{\mathrm{env}} = 14.0\% \pm 1.4\%$. The blue solid region indicates the 2$\sigma$ contours of the posterior distribution of the planet's reduced tidal quality factor, $Q'$, and $f_{\mathrm{env}}$ when accounting for tides. Bottom panel: The degree of radius inflation, parameterized as the ratio of the radius of the planet when tides are accounted for compared to the radius the planet would have if tides were absent. The blue solid region indicates the 2$\sigma$ contours of the posterior distribution. $Q'$ is highly uncertain but most likely in the range $Q' \approx 10^4 - 10^5$, indicating tidal inflation by a factor of $\sim1.5-1.7$.}
\label{fig:tidal inflation}
\end{figure}

As an illustrative example, we can estimate how much GJ 3470b's atmosphere has been inflated from tidal heating. According to traditional equilibrium tide theory \citep[e.g.,][]{Hut1981, Leconte2010}, the tidal luminosity---the rate of tidal energy dissipation inside the planet---is approximately $L_{\mathrm{tide}}\approx4.8\times10^{25}$ erg/sec, or roughly 4\% of the incident stellar power. Here we used the equations of \cite{Leconte2010} assuming zero planetary obliquity and a reduced tidal quality factor, $Q'$, equal to $10^5$. To quantitatively assess how this tidal luminosity affects the planetary structure, we fit a tidal model utilizing previous simulation results from \cite{Millholland2019} and \cite{Millholland2020}. The model consists of a spherically symmetric, two-layer planet containing a heavy element core and a H/He envelope. The thermal evolution of the atmospheric envelope is calculated using the Modules for Experiments in Stellar Astrophysics (MESA; \citealt{Paxton2011, Paxton2013}) 1D stellar evolution code, including a series of modifications developed by \cite{Chen2016} that make the model specific to planets as opposed to stars. Additionally, the model accounts for tidal energy deposited in the planet's deep atmosphere. The full details of the model are provided in \cite{Millholland2019} and \cite{Millholland2020}. Here we use an interpolation to the simulation set developed in \cite{Millholland2020}, and we use the Markov Chain Monte Carlo fitting procedure described therein to estimate the planet's envelope mass fraction ($f_{\mathrm{env}} \equiv M_{\mathrm{env}}/M_p$) and degree of tidal radius inflation. We assume that the eccentricity is fixed to $e=0.125$ (Table \ref{tab:rvparams}) and that the planetary obliquity is zero.

Figure \ref{fig:tidal inflation} shows the results of the tidal inflation analysis. The top panel indicates that the estimate of the planet's envelope mass fraction is much smaller when tides are included in the model compared to when they are ignored. This is because planets with active tidal heating are larger at a fixed $f_{\mathrm{env}}$, so the inclusion of tidal heating in the structural model yields smaller $f_{\mathrm{env}}$ estimates for observed planets. Moreover, smaller values of the planet's reduced tidal quality factor are associated with smaller values of $f_{\mathrm{env}}$, since the stronger dissipation at smaller $Q'$ requires a smaller envelope fraction to match the planet's observed radius. The bottom panel indicates that GJ 3470b is anywhere between $\sim1.2-1.7$ times larger than it would be in the absence of tides. However, a tidal quality factor in the range of $Q' \approx 10^4-10^5$ is most reasonable for this size of planet (\citealt{Millholland2020} and references therein). Thus, tidal heating has most likely inflated GJ 3470b by a factor of $\sim 1.5-1.7$.

With an inflated atmosphere, GJ 3470b is more susceptible to atmospheric escape than it would be in the absence of tidal heating. Tidal inflation, coupled with photoevaporation, thus serves as a potential connection between the orbital properties of the growing population of polar, eccentric Neptunes and their ongoing mass loss. We suggest future, more detailed studies on this connection.

\subsection{Possible Origin of the Polar Orbit}
\label{sec:formation}
How did GJ 3470b obtain its polar orbit? Different theoretical models have been developed to explain the origin of polar planetary orbits, and here we consider a few possible explanations. 

Highly inclined orbits can be produced through a primordial tilt of the protoplanetary disk \citep{batygin2012}, requiring interactions with a massive binary stellar companion or a stellar fly-by event \citep{malmberg2011}. Although we do not know if a stellar fly-by occurred in the system, GJ 3470 is not currently known to be in a stellar binary system. Additionally, polar orbits can be obtained through magnetic disk-torquing from the young magnetized star \citep{lai2011}. Although a plausible path to misaligning the disk and the orbit of GJ 3470b, this scenario acting alone could not easily explain the non-circular orbit of the planet ($e=0.122_{-0.041}^{+0.042}$).

Other theoretical scenarios that can explain the moderate eccentricity and polar orbit of GJ 3470b demand multi-body dynamical interactions. Among such models is planet-planet scattering \citep{rasio1996,chatterjee2008}. However, given the current large orbital velocity of GJ 3470b relative to its current escape speed ($[V_{\rm esc}/V_{\rm orb}]^2=0.053$), in-situ scattering is inefficient \citep{ford2001,petrovich2014}, favoring scattering at wider orbits. A possible formation scenario could then involve tidal high-eccentricity migration driven by planet-planet scattering, possibly accompanied by secular interactions via the Von Zeipel-Lidov-Kozai (ZLK) mechanism \citep[e.g.,][]{nagasawa2008}. Such models have been shown to be capable of explaining eccentric and misaligned orbits \citep[see e.g.,][]{bourrier2018,correia2020}, and recently specifically highlighted as a path to creating eccentric polar orbits by \cite{dawson2021}. Additionally, \cite{petrovich2020} presented a model relying on secular interactions between a massive outer planet and a slowly dissipating protoplanetary disk that is capable of explaining polar and eccentric orbits---and especially so for eccentric warm Neptunes orbiting around slowly rotating cool stars.

Given the tentative detection of an RV slope in Section \ref{sec:rv}, models that invoke secular dynamical interactions with an outer perturber are appealing. Below, we study two different secular excitation models: interactions from an inclined companion within the ZKL mechanism, and the disk-driven model of \cite{petrovich2020}. We show that both models are capable of explaining the polar orbit of GJ 3470b, and are compatible with the candidate RV slope for certain combinations of orbital distances and masses.

\begin{figure*}[t!]
\begin{center}
\includegraphics[width=0.9\textwidth]{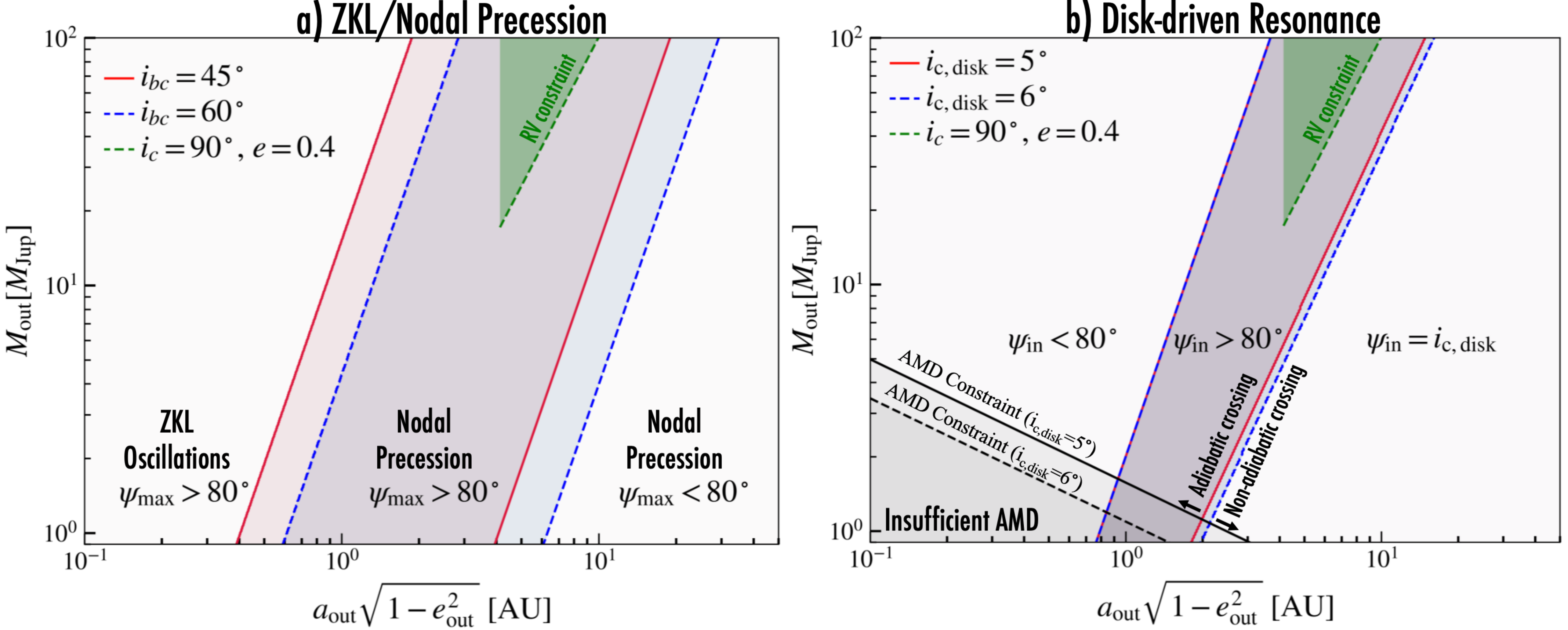}
\vspace{-0.5cm}
\end{center}
\caption{Constraints on the parameters of a potential outer perturber for the two models considered: a) the ZKL/nodal precession model, and b) the disk-driven model of \cite{petrovich2020}. For the former, the red and blue regions show where $\psi_{\rm max} > 80^\circ$ for $i_{\rm bc}=45^\circ$ and $i_{\rm bc}=60^\circ$, respectively. For the ZKL/nodal precession case, we assume a current Love number of $k_2=0.01$, leading to $J_2\sim 2\times 10^8$. For the disk-driven model of \cite{petrovich2020}, we assume that GJ 3470 had a radius of $R_\star = 0.9 R_\odot$ and rotation period of $10 \unit{days}$ and Love number of $k_2=0.2$ during the disk dispersal phase. For this case, we follow \cite{petrovich2020} and assume a disk mass-loss law of $M_{\mathrm{disk}} = 50M_{\rm J}/(1+t/\mathrm{1 Myr})^{1/2}$. The grey regions in b) highlight regions excluded from the conservation of angular momentum deficit (AMD). The green regions show the nominal region consistent with the RV slope from Section \ref{sec:rv}, where we have assumed $i_c \leq 90^\circ$, $e=0.4$ (the green dashed line shows the $i_c=90^\circ$ case).}
\label{fig:outer}
\end{figure*}

\subsection{ZKL Mechanism}
Within the ZKL mechanism, a highly inclined companion 'c' would change the orbital elements of planet b, including its nodal precession, on a timescale
\begin{equation}
\tau_{\mathrm{ZKL}} = \frac{2P_b}{3\pi }\left(\frac{b_c}{a_b}\right)^3 \frac{M_\star}{m_c}, 
\end{equation}
which is known as the ZKL timescale \citep[see][]{kiseleva1998}. Here, $b_c=a_c(1-e_c^2)^{1/2}$ is the semi-minor axis of the outer companion. This assumes that planet c has a high mutual inclination, which it might have obtained
through planet-planet scattering. Due to the close-in orbit of GJ 3470b, general relativistic (GR) precession and the rotationally-induced quadrupole of the host (especially early on its history) may modify the behavior of the ZKL oscillations. Following the notation of \citet{petrovich2020}, we define 
\ba
\label{eq:eta_gr}
\eta_{\rm GR}=\frac{8GM _\star}{ c^2} \frac{b_c^3}{a_b^4}\frac{M _\star}{m_c} \simeq 1\left(\frac{1M_{\rm J}}{m_c}\right)\left(\frac{b_c}{0.4\mbox{AU}}\right)^{3}
\ea
and 
\ba
\label{eq:eta_star}
\eta_\star=
\frac{2J_2M_\star}{m_c}\frac{R _\star^2b_c^3}{a_b^5}
\simeq 1\left(\frac{J_2}{10^{-6}}\right)
\left(\frac{1M_{\rm J}}{m_c}\right)\left(\frac{b_c}{2\mbox{AU}}\right)^{3}
\ea
where $J_2$ is the star's second zonal harmonic. These equations quantify the relative strength of GR corrections and the stellar quadrupole with respect to the two-planet interaction.

The general ZKL mechanism is capable of exciting both eccentricities and inclinations. An accurate treatment of the ZKL mechanism during periods of high-eccentricity oscillations could lead to orbital migration of the inner planet through tidal star-planet interactions during times of closest approach. For the purposes of this paper, we defer an analysis of star-planet tidal interactions to future work. Instead, we highlight the simpler and instructive special case of the ZKL mechanism in the limit where the eccentricity oscillations are quenched by general relativistic precession, resulting in only nodal precession of the inner planet's orbit. \cite{yee2018} described this special case of the ZKL mechanism as a possible explanation for the polar orbit of HAT-P-11b, which has a confirmed outer companion HAT-P-11c. As discussed by \cite{fabrycky2007}, eccentricity oscillations are quenched when
\begin{equation}
\sin i_{\rm bc} < \left(\frac{4+\eta_{\rm GR}}{10}\right)^{1/2},
\end{equation}
where $i_{\rm bc}$ is the mutual inclination between the planets' orbits and we have assumed that $\eta_{\rm GR} \gg \eta_\star$. The competition between the star's second zonal harmonic ($J_2$) and the possible outer planet orbit defines an equilibrium plane---the so-called \textit{Laplace plane}---inclined by $i_{\rm eq} (\eta_\star)$ relative the stellar equator (e.g., \citealt{laplace_surf}). Assuming the inner planet starts with zero obliquity, it will precess around the normal to this plane, sweeping out a cone with angle $i_{\rm eq}$ and driving oscillations of the obliquities from 0 to $\psi_{\mathrm{max}}=2i_{\rm eq}$, where
\ba
\psi_{\rm max}=\tan^{-1}\left[\frac{\sin(2i_{\rm bc})}{\cos(2i_{\rm bc})+\eta_\star}\right].
\ea
As expected, $\psi_{\rm max}=0$ when $\eta_\star \gg 1$ ($J_2$-dominated) and $\psi_{\rm max}=2i_{\rm bc}$ for $\eta_\star \ll 1$ (companion-dominated), passing through $\psi_{\rm max}=i_{\rm bc}$ at $\eta_\star=1$. 

In Figure \ref{fig:outer}a, we show the allowed properties of the outer planet that lead to $\psi_{\rm max}>80^{\circ}$ for two different choices of $i_{\rm bc}$, $45^\circ$ and $60^\circ$. From Figure \ref{fig:outer}, we see that the outer companion needs a semi-minor axis $b_{\mathrm{out}} \lesssim 10 \unit{AU}$ to result in $\psi_{\mathrm{max}}>80^\circ$. Further, Figure \ref{fig:outer}a shows that orbits with semi-minor axes between 1 and 10AU experience only nodal precession with no eccentricity oscillations for the inner planet. We also see that for closer-in orbits of $\lesssim1$AU, such orbits are capable of creating polar orbits for the inner planet but would excite both inclinations and eccentricities of the inner planet, likely leading to orbital migration of the inner planet. 

\subsection{Disk-Driven Resonance Model}
The model of \cite{petrovich2020} relies on secular interactions between an outer planet and a slowly dissipating protoplanetary disk, allowing for a process of inclination resonance sweeping and capture. In this model the final obliquity becomes
\ba
\label{eq:Icrit}
\psi=\sin^{-1}\left( \frac{4+4\eta _\star+\eta_{\rm GR}}{10+5\eta _\star}\right)^{1/2},
\ea
for $\eta_{\rm GR}<6+\eta _\star$ and $\psi=90^\circ$ for $\eta_{\rm GR}>6+\eta _\star$. The conditions for the resonance capture are:
\begin{enumerate}
\item The presence of a massive disk such that the resonance is encountered (condition in Eq. 2 in \citealt{petrovich2020}),
\item The process is adiabatic. This criterion requires the disk dispersal timescale $\tau_{\rm disk}=|d\log M_{\rm disk}/dt|^{-1}$ at crossing to be longer than the adiabatic timescale $\tau_{\rm ad}=\tau_{\rm ZKL}(1+\eta_\star)^{1/3}/i_{\rm c,disk}^{4/3}$, where $i_{\rm c,disk}$ is the inclination of the outer planet relative to the disk.
\end{enumerate}

Figure \ref{fig:outer}b graphically shows where the disk-driven model scenario of \cite{petrovich2020} suggests that polar orbits of GJ 3470b can be obtained. Similarly to the ZKL/nodal precession case in Figure \ref{fig:outer}a, we see that polar orbits of GJ 3470b can be obtained if a massive perturber (a few Jupiter masses) exists in the system at an orbital distance of $\sim$1-10AU. For the input parameters considered, we note that the disk-driven model case suggests a narrower parameter space where polar orbits can be produced. Similarly to the ZKL/Nodal precession case, we see from Figure \ref{fig:outer}b that the disk-driven model is also fully consistent with the RV slope (green region) discussed in Section \ref{sec:rv}.

Lastly, we note that the two models differ in their predictions for the mutual inclination between the inner planet and the outer companion. The ZKL/Nodal Precession case suggests a likely mutual inclination of $\sim$$45-60^\circ$, whereas the disk-driven resonance scenario suggests a mutual inclination closer to $90^\circ$. Future observations that constrain the orbital properties of the outer planet along with the mutual inclinations between the two planets can help distinguish between the two scenarios. 

\subsection{Future Constraints on a Possible Outer Companion}
A distant outer companion GJ 3470c with the right properties could explain the eccentric and polar orbit of GJ 3470b. Similar explanations were proposed for the polar orbits of GJ 436b, HAT-P-11b, and WASP-107b, the latter two of which have confirmed massive distant companions (see \citealt{yee2018}, and \citealt{piaulet2021}, respectively). An outer companion in the GJ 436 system has not yet been found (see \citealt{bourrier2018} for observational and theoretical constraints on potential outer companions). The existence of an outer companion GJ 3470c can be further confirmed with additional precise RVs, astrometric measurements from \textit{Gaia}, and direct imaging observations. 

Additional precise radial velocities will be immediately useful in confirming or ruling out the RV slope discussed above, and to see if such a slope starts to curve and reveal a periodicity. From both theoretical models discussed above, to produce polar orbits for the inner planet, we would expect the outer companion to be at a distance of a few AU. If the candidate RV slope discussed in Section \ref{sec:rv} is due to an acceleration from an outer companion, Figure \ref{fig:outer} shows that both model scenarios considered in the previous subsection are compatible with the slope. If the RV slope is not due to an acceleration due to an outer companion (e.g., stellar activity or other effects) there remains a possibility that a companion could be present in the system but at low inclinations that would not have been detected in the RV datasets.

Additionally, direct imaging would be sensitive to the most massive and distant outer perturbers within the regions in Figure \ref{fig:outer}. If we assume an orbital distance of $5 \unit{AU}$ (within the green region in Figure \ref{fig:outer}) the maximum sky-projected planet-star separation is $0.17 \unit{mas}$. Assuming a radius and albedo similar to Jupiter, the reflected-light contrast ratio of the planet is $\sim$$1.5\times 10^{-9}$, and could be a potential but challenging target for future constraints via high-contrast imaging. The Roman Space Telescope is expected to have a $10^{-9}$ effective contrast and $0.140\arcsec$ inner working angle\footnote{See \cite{spergel2015} and updated expected contrast numbers here: \url{https://github.com/nasavbailey/DI-flux-ratio-plot}.}, and may be able to characterize GJ 3470c if it exists. Additionally, we note that JWST MIRI, with an inner working angle of $0.33\arcsec$, could potentially place constraints on orbits beyond 10AU \citep[e.g.,][]{brande2020}.

Lastly, precise astrometric data from \textit{Gaia} can help constrain the possibility of an outer companion. We note that the \textit{Gaia} EDR3 astromety \citep{lindegren2021} shows that GJ 3470 has an excess astrometric noise of $0.18 \unit{mas}$ with an astrometric excess noise significance of $48\sigma$, which can be a sign of binarity. However, redder stars can also show significant astrometric noise independent of binarity \citep[e.g.,][]{thao2020}. The \textit{Gaia} Renormalized Unit Weight Error (RUWE) accounts for this color effect and has been shown to be a reliable indicator of binarity \citep[see e.g.,][]{ziegler2020}, where RUWE=1.0 indicates an astrometric solution consistent with a single star, with RUWE values larger than 1 indicative of non-single or extended sources. GJ 3470 has a RUWE=1.14. This value is consistent with a single M-dwarf star \citep[see e.g.,][]{thao2020} disfavoring a stellar binary, but we can not rule out the possibility of non-stellar companion in the system. Future \textit{Gaia} releases with access to the intermediate astrometric data products will enable more direct tests to probe for outer companions in the \textit{Gaia} data. In particular, \cite{sozzetti2014} calculated the expected fraction of M-dwarfs within 30\,pc for which \textit{Gaia} could detect a giant planet, finding an overall detection efficiency of $\sim$60\%, highlighting the possibility that an outer companion could be within the detectability threshold of the \textit{Gaia} mission.

\section{Conclusion and Summary}
\label{sec:summary}
By observing two transits with the newly commissioned NEID Spectrograph on the WIYN 3.5m Telescope, we showed that the warm Neptune GJ\,3470b has a polar orbit with a true obliquity of \respsi. We show that a well-aligned model ($\psi=0^\circ$) is strongly disfavored ($\Delta$BIC=58.6) relative to the best-fit misaligned model. This determination was facilitated by an improvement in the transit ephemeris using diffuser-assisted photometry with the ARC 3.5m Telescope at Apache Point Observatory.

GJ\,3470b joins a growing sample of warm Neptunes with nearly polar and mildly eccentric orbits, which could hint at the presence of a class of planetary systems that could share common formation histories. Using a tidal inflation model, we show that tidal heating due to GJ 3470b's mild eccentricity of $e\sim0.125$ has likely inflated the planet's radius by a factor of $1.5-1.7$, which can help account for its evaporating atmosphere. The polar and eccentric orbit of GJ 3470b---along with its evaporating atmosphere---together point to a formation scenario involving multi-body dynamical interactions, which likely includes interactions with a massive distant perturber. Using out-of-transit radial velocities spanning 13 years from HARPS, HIRES and HPF, we show that the RV data are compatible with a long-term RV slope at the $\sim$$2\sigma$ level although additional RV observations are needed to confirm or rule out this possible RV slope. Using two different secular excitation models, we constrain the possible orbital locations of an outer companion in the system, and we show that both models are compatible with the candidate RV slope. Future observations could both constrain the presence of an outer companion and further help distinguish between the two different secular excitation models through the measurement of the mutual inclinations between the inner planet and the potential outer companion.

\acknowledgements

We thank the anonymous referee for their thoughtful reading and suggestions, which made for a stronger manuscript. GKS thanks Luke Bouma and Molly Kosiarek for helpful discussions. Data presented were obtained by the NEID spectrograph built by Penn State University and operated at the WIYN Observatory by NOIRLab, under the NN-EXPLORE partnership of the National Aeronautics and Space Administration and the National Science Foundation. Based in part on observations at the Kitt Peak National Observatory, NSF’s NOIRLab (Prop. ID 2020B-0075; PI: G. Stefansson), managed by the Association of Universities for Research in Astronomy (AURA) under a cooperative agreement with the National Science Foundation. WIYN is a joint facility of the University of Wisconsin–Madison, Indiana University, NSF’s NOIRLab, the Pennsylvania State University, Purdue University, University of California, Irvine, and the University of Missouri. The authors are honored to be permitted to conduct astronomical research on Iolkam Du’ag (Kitt Peak), a mountain with particular significance to the Tohono O’odham. Data presented herein were obtained at the WIYN Observatory from telescope time allocated to NN-EXPLORE through the scientific partnership of the National Aeronautics and Space Administration, the National Science Foundation, and the National Optical Astronomy Observatory. C.P. acknowledges support from ANID Millennium Science Initiative-ICN12\_009, CATA-Basal AFB-170002, ANID BASAL project FB210003, FONDECYT Regular grant 1210425 and ANID+REC Convocatoria Nacional subvencion a la instalacion en la Academia convocatoria 2020 PAI77200076. 

These results are based on observations obtained with the Apache Point Observatory 3.5-meter telescope which is owned and operated by the Astrophysical Research Consortium. We wish to thank the APO 3.5m telescope operators in their assistance in obtaining these data.

This work was partially supported by funding from the Center for Exoplanets and Habitable Worlds. The Center for Exoplanets and Habitable Worlds is supported by the Pennsylvania State University, the Eberly College of Science, and the Pennsylvania Space Grant Consortium. CIC acknowledges support by NASA Headquarters under the NASA Earth and Space Science Fellowship Program through grants 80NSSC18K1114. This work was performed for the Jet Propulsion Laboratory, California Institute of Technology, sponsored by the United States Government under the Prime Contract 80NM0018D0004 between Caltech and NASA. We acknowledge support from NSF grant AST-1909506, AST-190950, AST-1910954, AST-1907622 and the Research Corporation for precision photometric observations with diffuser-assisted photometry. Computations for this research were performed on the Pennsylvania State University’s Institute for Computational \& Data Sciences (ICDS). A portion of this work was enabled by support from the Mt Cuba Astronomical Foundation. 

These results are based on observations obtained with the Habitable-zone Planet Finder Spectrograph on the HET. We acknowledge support from NSF grants AST 1006676, AST 1126413, AST 1310875, AST 1310885, and the NASA Astrobiology Institute (NNA09DA76A) in our pursuit of precision radial velocities in the NIR. We acknowledge support from the Heising-Simons Foundation via grant 2017-0494. This research was conducted in part under NSF grants AST-2108493, AST-2108512, AST-2108569, and AST-2108801 in support of the HPF Guaranteed Time Observations survey. The Hobby-Eberly Telescope is a joint project of the University of Texas at Austin, the Pennsylvania State University, Ludwig-Maximilians-Universitat Munchen, and Georg-August Universitat Gottingen. The HET is named in honor of its principal benefactors, William P. Hobby and Robert E. Eberly. The HET collaboration acknowledges the support and resources from the Texas Advanced Computing Center. We thank the Resident astronomers and Telescope Operators at the HET for the skillful execution of our observations with HPF.

This work is based on observations made with the HARPS spectrograph on the 3.6m ESO telescope at the ESO La Silla Observatory, Chile, under programs 082.C-0718(B), 183.C-0437(A), 089.C-0050(A), and 198.C-0838(A) publicly available through the ESO archive (\url{http://archive.eso.org/eso/eso_archive_main.html}).

This work has made use of data from the European Space Agency (ESA) mission Gaia processed by the Gaia Data Processing and Analysis Consortium (DPAC). Funding for the DPAC has been provided by national institutions, in particular the institutions participating in the Gaia Multilateral Agreement.

This research made use of the NASA Exoplanet Archive, which is operated by the California Institute of Technology, under contract with the National Aeronautics and Space Administration under the Exoplanet Exploration Program. This research made use of Astropy, a community-developed core Python package for Astronomy \citep{astropy2013}.


\facilities{NEID/WIYN 3.5m, ARCTIC/ARC 3.5m, HPF/HET 10m, HIRES/Keck 10m, HARPS/La Silla 3.6m \textit{Gaia}} 
\software{AstroImageJ \citep{collins2017}, 
\texttt{astroplan} \citep{morris2018},
\texttt{astropy} \citep{astropy2013},
\texttt{astroquery} \citep{astroquery},
\texttt{barycorrpy} \citep{kanodia2018}, 
\texttt{batman} \citep{kreidberg2015},
\texttt{celerite} \citep{Foreman-Mackey2017}, 
\texttt{corner.py} \citep{dfm2016}, 
\texttt{dynesty} \citep{speagle2019}, 
\texttt{emcee} \citep{dfm2013},
\texttt{iDiffuse} \citep{stefansson2018b},
\texttt{juliet} \citep{Espinoza2019},
\texttt{Jupyter} \citep{jupyter2016},
\texttt{matplotlib} \citep{hunter2007},
\texttt{numpy} \citep{vanderwalt2011},
\texttt{pandas} \citep{pandas2010},
\texttt{pyde} \citep{pyde},
\texttt{radvel} \citep{fulton2018},
\texttt{SERVAL} \citep{zechmeister2018}.}

\bibliography{references}{}
\bibliographystyle{aasjournal}

\end{document}